\documentstyle[aps,epsf,twocolumn]{revtex}

\begin{document}
\draft

\twocolumn[\hsize\textwidth\columnwidth\hsize\csname @twocolumnfalse\endcsname

\title{Electrostatics of Lipid Bilayer Bending}

\author{Tom Chou, Marko V. Jari\'{c} \footnote{Permanent
Address: Center for Theoretical Physics, Texas A\&M
University, College Station, TX 77843-4242.}, 
and Eric D. Siggia}

\address{Laboratory of Atomic and Solid State Physics,
Clark Hall, Cornell University, Ithaca, NY 14853}

\date{\today}
\maketitle

\begin{abstract}
The electrostatic contribution to spontaneous 
membrane curvature is calculated within
Poisson-Boltzmann theory under a variety of 
assumptions and emphasizing parameters in 
the physiological range. Asymmetric surface charges, 
either fixed with respect to bilayer midplane area, 
or with respect to the lipid-water area both 
induce curvature but of opposite sign.  
Unequal screening layers on the two sides of a
vesicle ({\it e.g.} multivalent cationic proteins on one
side and monovalent salt on the other) 
also induces bending. For reasonable parameters,
tubules formed by electrostatically induced 
bending can have radii 
in the 50-100nm range, often seen in many intracellular
organelles. Thus membrane associated proteins
may induce curvature and subsequent budding, 
without themselves being intrinsically curved. 
Furthermore, we derive the previously unexplored 
effects of respecting the strict conservation of
charge within the interior of a vesicle. 
The electrostatic 
component to the bending modulus is small under
most of our conditions, and is left as an 
experimental parameter. The large parameter 
space of conditions is surveyed in an array of graphs.
\end{abstract}

\bigskip
\noindent Keywords: membrane curvature, 
bilayer electrostatics, vesicles, 
Poisson-Boltzmann equation. 
\bigskip
]

\vspace{8mm}

\noindent {\bf INTRODUCTION}

\noindent The membranes bounding intracellular organelles 
are dynamic structures,
whose morphology the cell can regulate.  
This happens most dramatically
during mitosis, when the nuclear membrane 
disintegrates, probably forming vesicles 
(Alberts, 1994). Proteins destined for 
secretion or targeted to the plasma membrane pass
from the endoplasmic reticulum (ER) to the 
Golgi body and thence to their target; at each step 
vesicles or tubular processes with diameters 
in the 50-100nm range are involved in the 
sorting and transport (Rothman, 1994;
Schekman and Orci, 1996). Similar remarks 
apply to endocytosis and endosomal sorting 
and maturation (Gruenberg and Maxfield, 1995; 
Trowbridge, et al., 1993). The ER cisterna itself 
has a large tubular component (Terasaki et al., 1986) 
and there are tubular connections  within the 
Golgi body (Rambourg and Clermont, 1990). Various 
treatments can enhance the tubulation of 
the membranes of Golgi and other organelles
(Lippincott-Schwartz et al., 1991; Cluett et al., 1993).

Numerous plausible mechanisms have been 
proposed for vesicle budding or tubule formation,
but often there is insufficient physical detail 
for a quantitative assessment of their validity.
For instance, are the building blocks of the 
clathrin cages intrinsically curved,
or do kinetic processes during budding favor 
five membered over six membered rings and 
thus effect closure (Shraiman, 1996).  
The phase separation of wedge shaped lipids or membrane 
resident proteins, is also argued to play a role
in vesiculation (Schekman and Orci, 1996), as is 
line tension between the two phases (Lipowsky, 1993).
Chemical modifications to the lipids on one 
side of a bilayer, such as phosphorylation of 
inositol lipids (de Camilli et al., 1996), 
or cleavage of
the acyl chains by phospholipases (Brown, 1996), 
are known to occur biologically and will promote 
membrane curvature.  Less drastic
modifications such as changes in solution pH 
or ionic strengths
can have similar effects because many lipids 
are zwitterionic and their effective 
charge varies with solution conditions. 
Furthermore, since parameters such as pH, 
ionic strength, temperature, phospholipid 
pK$_{a}$ are not independent, 
it is impossible to separate and quantify 
all these effects.

It has long been recognized that when surface charges
are unequal, electrostatics will induce membrane
curvature. In this Article we quantify a number of
less obvious influences of electrostatics on membrane
bending.  Symmetrically charged 
membranes (i.e., same
surface charge on the two leaflets) will bend
in response to asymmetrical screening.  Thus membrane
associated proteins such as the 
adaptins which mediate between the 
clathrin and the bilayer
could cause bending, even if the proteins are globular
while free in solution. Various plausible choices
for the neutral surface with respect 
to which the surface charge 
is conserved during membrane bending can alter even
the sign of the preferred curvature resulting from a
given charge or screening asymmetry. 
Uncharged membrane lipid components can define the
neutral surface and hence modulate the electrostatic
response. Finally, if the interior of a vesicle is
truly electrically isolated from the exterior, a small 
internal charge scaling with the area can define the 
spontaneous curvature. The simple description of
electrostatics we employ, the Poisson-Boltzmann
equation, has been shown to give quantitatively
accurate results for the arguably more complex problem
of the binding energy between a charged protein and a
bilayer (Tal {\it et al.} 1996).

Even at the level of Poisson-Boltzmann
theory, the electrostatic contribution to 
spontaneous curvature particularly with
asymmetrical (different on the two sides) 
and multivalent electrolytes have not been 
fully quantified in the literature, and are
potentially significant for the concentrated 
electrolytes encountered in the cell.  
Our interest was kindled by the ubiquity of tubules
within the diameter range noted above, and
the possibility of getting such sizes by 
simple electrostatics. More attention has 
been paid in the physical literature to the
electrostatic contribution to the curvature 
energy (we will generally take the total bending modulus
as an experimentally determined parameter)
rather than the spontaneous curvature. 
Low ionic strength (which enhances the 
electrostatic contribution to the bending modulus) 
was emphasized when the spontaneous curvature
was computed, and symmetrical, typically 
monovalent electrolytes were assumed (Mitchell and
Ninham 1989; Winterhalter and 
Helfrich, 1989; 1992).

Our calculations are limited to the equilibrium 
radius of a sphere or tube (with ends neglected), 
and we ignore the more subtle problem of 
determining the overall shape of a homogeneous 
vesicle with given volume, area, and leaflet
area difference (a surrogate for the spontaneous 
curvature) (Seifert and Lipowsky 1995; Mui 1995). 
Biological systems are very 
inhomogeneous; present are channels
that span the membrane, and enzymes that circumvent
kinetic barriers to lipid repartitioning that 
are a feature of {\it in vitro} systems.  
In short, the electrostatic effects we are considering
may be the physical adjunct to some of the biochemically
defined actors, (wedge shaped lipids, adaptins, 
coatamers etc.), involved in vesicle and tubule 
formation.

\begin{figure}[htb]
\begin{center}
\leavevmode
\epsfysize=3.5in
\epsfbox{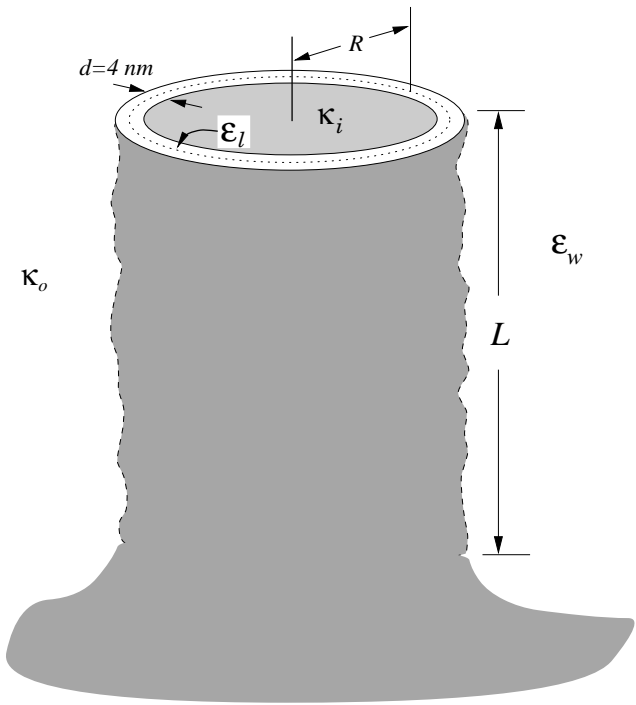}
\end{center}
\caption{Schematic of a tubular section of
a vesicle. The structure is assumed cylindrical 
and has radius $R$ and thickness $d$. The interior and
exterior screening lengths are $\kappa_{i}^{-1}$ and
$\kappa_{o}^{-1}$, respectively. Our 
calculations assume $L\gg R$. }
\label{FIG1}
\end{figure}

It should not be forgotten that electrostatic 
effects can be large in a biological context 
(Honig and Nicholls, 1995); for a surface charge of 
-0.2$\vert e\vert$/nm$^{2}$ (corresponding to 
10\% of lipids each of size $\sim 50$\AA$^{2}$ possessing
an electronic charge), a screening length of 1nm, and an
aqueous dielectric constant $\epsilon_w \simeq 80$, 
the surface potential $e\varphi \sim k_B T$, 
the thermal energy. Several parameters that occur repeatedly 
in this paper are diagrammed in Figure \ref{FIG1} 
for a single tubule of radius $R$ on an isolated 
vesicle.

In the following section, we recapitulate the nonlinear 
electrostatic free energy  (there is no unanimity 
among prior papers) and
collect several formulae for the linearized
free energy, and the leading $1/R$ term in the
nonlinear free energy. We emphasize the effects of
multivalent electrolytes and interleaflet coupling. 
The surface charge responds to the curvature 
under two possible scenarios; either not at all 
{\it i.e.,} fixed charge per midplane area;
or maximally, with charge fixed per area of each 
water-lipid interface (and separate reservoirs 
assumed for each leaflet). Equilibrium radii are very
different in the two cases.

We also take literally 
the isolation between the interior and exterior
electrolytes effected by the membrane of a giant 
lipid vesicle {\it in vitro} to motivate a calculation 
within a fixed number ensemble for the interior.
Theories which assume a thermodynamic reservoir ({\it e.g.}
Poisson-Boltzmann) are incorrect unless the total internal
charge is set to a particular value to within
$O(\epsilon_{\ell}/\epsilon_{w})$ times the total surface
charge. Deviations from this value can qualitatively change
the conclusions. In the Results,
we compute various quantities using realistic
parameters and contrast the results and various 
assumptions. The concluding section 
reviews pertinent experiments with a view
towards extracting values of parameters 
and supporting the various assumptions we have made.

\vspace{8mm}

\noindent {\bf ELECTROSTATICS}

\noindent In this study, we assume all structures are 
smooth on atomic length scales and adopt 
the continuum limit. The problem is that of a
cylindrical or spherical shell of thickness 
$b-a = d = 4$nm and 
dielectric constant $\epsilon_{\ell} = 2$ 
embedded in an ionic solution with dielectric
constant $\epsilon_{w} = 80$. We assume that 
changing the aqueous buffer ionic strength 
does not affect $\epsilon_{w}$. These parameters 
approximate a lipid bilayer membrane under 
{\it in vitro} and {\it in vivo} conditions. 

The total charge density $\rho(\vec{r})$ of the 
system can be decomposed into
an aqueous mobile  contribution and a fixed piece,

\begin{equation}
\rho(\vec{r}) = \sum_{i}\rho_{i, fixed} +
\sum_{i}\rho_{i, mobile},
\end{equation}

\noindent where the fixed charge contribution
resides exclusively at the lipid-water interface
and is characterized by a surface charge $\sigma$. 
The surface charge results from 
ionization of {\it e.g.} phosphate, amine, and
carboxyl groups on hydrophilic moieties of the
lipids; these effective surface charges are 
typically negative. We do not consider 
spatially varying surface charges and lateral lipid phase
separation; however, for dilute charge densities, 
and smaller screening lengths, the 
interactions among charged lipids 
are expected to be minimal (Hui, 1996) and the charged 
component would be approximately a uniform 2D ideal gas
over the entire vesicle. Details of how $\sigma$
is affected by solution conditions, local membrane
curvature, etc. are deferred until after results are 
presented.

The mobile charge density is assumed 
to follow a Boltzmann ensemble

\begin{equation}
\rho_{i, mobile}(\vec{r}) = \rho_{i}^{\infty}
e^{-e\beta\varphi(\vec{r})}.
\end{equation}

\noindent where $\rho_{i}^{\infty}$ is the 
bulk concentration of specie $i$ and $\beta^{-1} \equiv
k_{B}T$. Substituting $\rho_{mobile}(\vec{r})$ into 
Poisson's equation,

\begin{equation}
\nabla\cdot(\epsilon(\vec{r})\nabla\varphi) = -4\pi
\rho(\vec{r}),
\label{POISSON}
\end{equation} 

\noindent we obtain, for the electrostatic 
potential in the aqueous phase, 

\begin{equation}
\begin{array}{r}
\displaystyle \ell^{2} \nabla^{2}\psi = 
n_{1}\sinh \psi +
n(z_{+}){z_{+}\over 2}\left(e^{\psi}-
e^{-z_{+}\psi}\right)+ \\[13pt]
\displaystyle {z_{-} \over
2}n(z_{-})\left(e^{z_{-}\psi}-e^{-\psi}\right), \quad
\label{DIFFEQ}
\end{array}
\end{equation}

\noindent where $e\beta \varphi \equiv \psi$, 
and $n_{1},\, n(z_{+})$, and $n(z_{-})$ are 
salt concentrations in moles/liter 
of monovalent, $z_{+}$-valent and
$z_{-}$-valent ({\it e.g.} for CaCl$_{2}$,
$z_{+} = 2$; for Na$_{2}$SO$_{4}$, $z_{-} = 2$)
ions respectively. Lengths are 
expressed in nanometers and all 
numerical conversion factors are 
subsumed in $\ell = 0.3083$nm (T=300K).

The electrostatic portion of the free energy for the lipid
plus buffer system follows most readily in the limit 
where Eq. \ref{DIFFEQ} is linearized. Thus if a uniform
surface charge is given,

\begin{equation}
G_{e\ell} = {1 \over 2}\sum_{i}\sigma_{i}\varphi_{i}
\end{equation}

\noindent where $i$ labels all surface charge positions.
This prescription can be ambiguous in the nonlinear case
(Sharp and Honig 1990).

An alternative argument is then to use the
expression whose variation yields Eq. \ref{DIFFEQ}
and which agrees with the linear theory in the 
appropriate limit (Dresner 1963; Sharp and Honig 1990). 
Thus, with fixed surface charge, 

\begin{equation}
\begin{array}{r}
\displaystyle G_{e\ell} = \int dS \sigma\varphi - 
{\epsilon_{w}\over 4\pi}
\int^{\prime}d^{3}r \left[{1 \over
2}\vert\nabla\varphi\vert^{2}+U\left[\varphi\right]
\right]- \\[13pt]
\displaystyle  {\epsilon_{\ell} \over 4\pi}
\int^{\prime\prime}d^{3}r {1\over
2}\vert\nabla \varphi\vert^{2}, \quad
\label{G0}
\end{array} 
\end{equation}

\noindent where the primed integral is taken only over
coordinates in the aqueous solution and the 
double-primed integral is taken over the 
bilayer region ($a<r<b$) occupied by the
lipid acyl chains. This equation is also the large volume
limit of a functional that was shown to be equivalent 
to the standard thermodynamic definition of 
$G_{e\ell}\left[\sigma\right]$ (Dresner 1963), as well as 
expressions involving a parameter integral (Marcus 1955).
We have checked that Eq. \ref{G0}
can be reduced to a form similar to the frequently
employed charging integral 
(Marcus, 1955; Sharp and Honig, 1990), only in the 
one-dimensional case and for all dimensions in 
linearized theory. Furthermore, we have explicitly
verified that the $1/R$ coefficient in an expansion
of $G_{e\ell}(R)/2\pi RL$ also agrees with that found from the
charging integral when membrane leaflets are 
uncoupled. 

The free energy with fixed surface potential is 
just Eq. \ref{G0} with the first term omitted, 
$G_{e\ell}\left[\varphi(S)\right] =
G_{e\ell}\left[\sigma\right]-
\int\varphi(S)\sigma dS$. In all cases, Eq. \ref{G0} is
understood to be evaluated for a stationary solution,
{\it i.e}, Eq. \ref{DIFFEQ}.

The ``potential'' $U\left[\varphi\right]$ is just 

\begin{equation}
\begin{array}{r}
\displaystyle (\ell e\beta)^{2}U\left[\varphi\right] 
\equiv n_{1}\left(\cosh \psi - 1\right)  \quad \quad
\quad \quad  \\[13pt]
\displaystyle + n(z_{+}){z_{+}\over 2}\left(e^{\psi}+
{e^{-z_{+}\psi} \over z_{+}}-1-
{1 \over z_{+}}\right)  \\[13pt]
\displaystyle + n(z_{-}){z_{-} \over 2}\left(e^{-\psi} +
{e^{z_{-}\psi} \over z_{-}}-1-{1 \over z_{-}}\right)
\end{array}
\end{equation}

\noindent in the units of Eq. \ref{DIFFEQ} and vanishes 
as $\varphi \rightarrow 0$ to make Eq. \ref{G0} scale with the
surface area. The surface value of $\varphi$ is also to be
varied and the net coefficient of $\delta\varphi(S)$ is 
the usual boundary condition 

\begin{equation}
\sigma + {\epsilon_{w}\over 4\pi}\partial_{n}
\varphi_{w} - {\epsilon_{\ell}\over 4\pi}
\partial_{n}\varphi_{\ell} = 0
\label{BC0}
\end{equation}

\noindent at all lipid-water interfaces. The normal
derivative is taken positive from the lipid outwards. 

\vspace{8mm}
\noindent {\bf Linear Poisson-Boltzmann Solutions}

\noindent For potentials $e\beta\varphi \ll 1$, 
the linearized form of the equation of motion
to be solved in the aqueous solution 
with the proper fixed charge boundary
conditions is 

\begin{equation}
\nabla^{2}\varphi - \kappa_{i,o}^{2}\varphi = 0 
\label{DIFFEQLIN}
\end{equation}

\noindent where $\kappa_{i,o}$ are the 
effective inverse screening lengths 
inside or outside the vesicle ($z_{\pm}>0$),

\begin{equation}
\ell^{2} \kappa^{2} = n_{1}+{1 \over
2}n(z_{+})z_{+}(z_{+}+1)
+{1 \over 2}n(z_{-})z_{-}(z_{-}+1).
\label{KAPPA}
\end{equation}

\noindent Inside the bilayer $(a<r<b)$, there are no
charges, $U = 0$, and Laplace's equation holds. 
The cylindrically symmetric solutions in all regions are
also explicitly displayed in Appendix A. The corresponding 
linearized form of the free energy (Eq. \ref{G0})
is $G_{e\ell} = {1 \over 2}\int dS \sigma\varphi$
and the free energy per midplane area,
$2\pi L R$ is

\begin{equation}
g_{e\ell} = {1\over 2}\sigma_{a}\varphi(a)
\left(1-{d \over 2R}\right)
+ {1 \over 2}\sigma_{b}\varphi(b)
\left(1+{d\over 2R}\right).
\label{GLIN}
\end{equation}
 
\noindent Expressions in the decoupled limit 
$\epsilon_{\ell}/(\epsilon_{w} \kappa d) = 0$ have 
previously been found (Winterhalter and
Helfrisch 1988). We will analyze consequences of the
expansion 

\begin{equation}
g_{e\ell} = C_{0}+{C_{1} \over R} + {C_{2} \over R^{2}}
+O\left({1\over R^{3}}\right)
\end{equation}

\noindent for screening lengths appropriate to 
physiological conditions.

Further expanding the coefficients 
$C_{i}$ in powers of $\epsilon_{\ell}/
(\epsilon_{w}\kappa d)$, we obtain

\begin{equation}
\begin{array}{r}
\displaystyle C_{0} = {2 \pi d \over
\epsilon_{w}}\left[\left({\sigma_{a}^{2}\over
\kappa_{i}d}+{\sigma_{b}^{2} \over \kappa_{o}d}\right)
+{\epsilon_{\ell}\over \epsilon_{w}}{(\kappa_{i}d
\sigma_{b}-\kappa_{o}d\sigma_{a})^2 \over
(\kappa_{i}d\kappa_{o}d)^2}\right] + \\[13pt]
\displaystyle O\left({\epsilon_{\ell} 
\over \epsilon_{w}\kappa d}
\right)^2
\label{C0}
\end{array}
\end{equation}

\begin{equation}
\begin{array}{r}
\displaystyle C_{1} = {\pi d^2 \over
\epsilon_{w}}\bigg[{\sigma_{b}^2\over
(\kappa_{o}d)^2}(\kappa_{o}d-1)-{\sigma_{a}^{2}\over
(\kappa_{i}d)^2}(\kappa_{i}d-1) + \\[13pt]
\displaystyle {\epsilon_{\ell}\over 
\epsilon_{w}\kappa_{i}^{3}
\kappa_{o}^{3}d^3}\bigg(\sigma_{a}^{2}
(\kappa_{i}^2 \kappa_{o}-\kappa_{i}\kappa_{o}^2 -
2\kappa_{o}^3)+ \quad \quad \\[13pt]
\displaystyle \quad \quad \sigma_{a}\sigma_{b}\kappa_{i}
\kappa_{o}(\kappa_{o}-\kappa_{i})+
2\sigma_{b}^2 \kappa_{o}^{3}\bigg)
+O\left({\epsilon_{\ell}\over \epsilon_{w}\kappa
d}\right)^{2}\bigg]
\label{C1}
\end{array} 
\end{equation}

\noindent and

\begin{equation}
C_{2} = 
{3\pi d^{3} \over 4\epsilon_{w}}\left[
{\sigma_{a}^{2} \over (\kappa_{i}d)^{3}}
+{\sigma_{b}^{2} \over (\kappa_{o}d)^{3}}
+ O\left({\epsilon_{\ell}\over \epsilon_{w}\kappa
d}\right)\right].
\label{C2}
\end{equation}

\noindent With this notation,
the total membrane bending stiffness is 
$2C_{2} + k_{m}$ where $k_{m}$ represents 
bending stiffness from nonelectrostatic (such as
mechanical) contributions.
When $C_{1}<0$ the membrane spontaneously
curves in the sense we have assumed ($b=$ out, $a=$ in)
and it is interesting to understanding the physical 
origin of the effect. For $\kappa d \ll 1$ either 
$\sigma_{b} > \sigma_{a}\, (\kappa_{i} 
\simeq \kappa_{o})$ or $\kappa_{o}^{-1} > 
\kappa_{i}^{-1}\, (\sigma_{b} \simeq \sigma_{a})$ 
favors bending. This prediction follows by
replacing each screening layer by cylindrical 
capacitors; the larger charge, or thicker 
layer will prefer the exterior, {\it i.e.} 
the screening charge cloud expands.
In the opposite limit ($\kappa d \gg 1$), 
we can think of the two thin 
screening layers as flat; here,
the energy is minimized by making the layer with the
greatest energy/area be interior, {\it i.e.} occupy the
side with less area.

The influence of $\epsilon_{\ell}$ is felt through the
dimensionless combination 
$\epsilon_{\ell}/(\epsilon_{w}\kappa
d)$ as expected from Eq. \ref{BC0}.
This is always small under our conditions. Note that it
is incorrect to estimate the energy as a triple 
of independent capacitors of thickness 
$\kappa_{i}^{-1}, d,$ and $\kappa_{o}^{-1}$, 
{\it i.e.} $\sigma^{2} d/\epsilon_{\ell} + 
O(\sigma^{2}/\kappa\epsilon_{w})$ which would 
make the lipid contribution appear dominant. 
Also note that for $\epsilon_{\ell}/\epsilon_{w}
\kappa d\rightarrow 0$, $C_{1}$ is 
antisymmetric with the interchange
$\sigma_{a} \leftrightarrow \sigma_{b}, \, \kappa_{i}
\leftrightarrow \kappa_{o}$. When coupling through the
bilayer is not negligible, then the antisymmetry exists
only for $\kappa_{i} = \kappa_{o}$ or $\sigma_{a} =
\sigma_{b}$.

Finally, we consider a scenario in which the surface
charges are fixed with respect not to their physical 
interface area, but with respect to an area defined by a radius
in the interior of the lipid. This radius defines 
a ``neutral surface'' (Petrov and Bivas 1984).
We consider here the extreme case 
of conserved charge per midplane area, $A = 2\pi R L$. 
The physical reasons for this choice as opposed to 
fixed charge per area of membrane/solution interface
is deferred to the Discussion. The only change to 
Eqs. \ref{C0}, \ref{C1}, and \ref{C2} is a redefinition 
of $\sigma_{a,b}$. In the decoupled limit, 
upon replacing 
$\sigma_{a,b} \rightarrow \sigma_{a,b}/
(1\pm d/2R)$, the
coefficients $C_{i}$ for $g_{e\ell}(R)$ 
become

\begin{equation}
C_{1} \simeq {\pi d^2 \over
\epsilon_{w}}\left[{\sigma_{a}^{2} \over
(\kappa_{i}d)^2}(1+\kappa_{i}d) - {\sigma_{b}^{2} 
\over (\kappa_{o}d)^{2}}(1+\kappa_{o}d)\right]
\label{CM1}
\end{equation}

\noindent and

\begin{equation}
\begin{array}{r}
\displaystyle C_{2} \simeq {\pi d^{3} \over 4\epsilon_{w}}
\bigg[\left({3 \over (\kappa_{i}d)^3} + {4 \over
(\kappa_{i}d)^2} + {2 \over
\kappa_{i}d}\right)\sigma_{a}^{2} + \quad \\[13pt]
\displaystyle \left({3 \over (\kappa_{o}d)^3} + {4 \over
(\kappa_{o}d)^2} + {2 \over
\kappa_{o}d}\right)\sigma_{b}^{2}\bigg]
\label{CM2}
\end{array}
\end{equation}

\noindent with $C_{0}$ remaining unchanged. Note in 
Eqs. \ref{CM1} and \ref{CM2} $\sigma_{a,b}$ are now the fixed
charges per midplane area. Although
$C_{2}$ changes only slightly in magnitude,
the behavior of $C_{1}$ 
is qualitatively different from Eq. \ref{C1}.

\vspace{8mm}
\noindent {\bf Expansion of Nonlinear Poisson-Boltzmann
Equation}

\noindent It is often noted (Mitchell and Ninham 1992;
Winterhalter and Helfrich, 1992) that the 
first few terms of an expansion of $G_{e\ell}$ in terms 
of $d/R$ or $1/(\kappa R)$
can be analytically obtained for the nonlinear theory.
The first order $(1/R)$, is particularly simple 
to derive because one is expanding a stationary 
functional about its minimum and several terms 
cancel. The equation to be solved is 

\begin{equation}
-\nabla^{2}\varphi + {\partial U \over \partial
\varphi} = -{\partial^2 \varphi \over \partial r^2}
-{1 \over r}{\partial \varphi \over \partial r} + 
{\partial U \over \partial
\varphi} = 0,
\label{DIFFEQ2}
\end{equation}

\noindent with the boundary conditions given by
Eq. \ref{BC0}.
Let $\varphi_{0}$ satisfy 
Eq. \ref{DIFFEQ2} without the 
$r^{-1}\partial_{r}\varphi(r)$ term. Upon using the
boundary condition $\varphi_{0}\rightarrow 0$ far from 
charged surfaces and $U(0) = 0$, this equation 
for $\varphi_{0}(r<a, r>b)$ can be integrated to yield

\begin{equation}
{1 \over 2}(\partial_{r}\varphi_{0})^2 - 
U\left[\varphi_{0}\right] = 0.
\label{DIFFEQ3}
\end{equation}

\noindent For negatively charged interfaces,
$\varphi_{0}(r=a,b)$ is negative and 
follows from Eq. \ref{DIFFEQ3} with the boundary
conditions $\lbrack\lbrack \epsilon(r)\nabla\varphi_{0}
\rbrack\rbrack = -4\pi\sigma$. 
Here, $\lbrack\lbrack \ldots\rbrack\rbrack$ denotes
the discontinuity across the charged surface.
The free energy per midplane area from Eq. \ref{G0} is
the sum $g_{e\ell}(R) = g_{+}(R)+g_{-}(R)+g_{\ell}(R)$
where $g_{-},\, g_{+},$ and $g_{\ell}$ are 
contributions from the inner, outer and bilayer portions
of the membrane system. 
One can readily show (Appendix B) 
that $g_{\ell}(R)$ contributes 
only even powers of $1/R$; the effect of bilayer 
coupling via $\epsilon_{\ell}$, however, is still felt
through the surface values of $\varphi_{0}$ and
$\partial_{r}\varphi_{0}$ determined from the boundary
conditions.  In Appendix B, 
we explicitly perform the calculation 
of $g_{e\ell}(R)$ to $O(1/R)$ and obtain

\begin{equation}
\begin{array}{r}
\displaystyle g_{e\ell} \simeq g(\infty) + 
{\epsilon_{w} \over 4\pi R}\int_{0}^{\infty} 
({\cal E}_{-}-{\cal E}_{+}) z\,dz + {d \over
2R}\bigg[\sigma_{b}\varphi_{0}(b)- \\[13pt]
\displaystyle \sigma_{a}\varphi_{0}(a) + {\epsilon_{w}
\over 4\pi}\int_{0}^{\infty}
({\cal E}_{-}-{\cal E}_{+})dz \bigg] \quad\quad
\label{gTOT}
\end{array}
\end{equation}

\noindent where all inner and outer
``one-dimensional'' energies ${\cal E}_{\pm}
\equiv {1 \over 2}(\partial_{r}\varphi)^2+
U\left[\varphi\right]$ are evaluated with
$\varphi = \varphi_{0}(z)$ and $\partial_{r} \rightarrow
\partial_{z}$. The second and third terms in 
Eq. \ref{gTOT} give the $1/R$ dependence of the 
free energy. We have also assumed that the inner 
radial surfaces of the tube ($r=a$) are far enough 
apart such that $a \gg
\kappa_{i}^{-1}$. 

\vspace{8mm}
\noindent {\bf Nonlinear Numerical Calculation}

\noindent For smaller $R$, higher order terms in $1/R$
must be taken into account or the 
full nonlinear solution for $\varphi(r)$ 
and the corresponding free energy
must be found numerically. The inner and outer
solutions are matched via the 
surface charge dependent discontinuous derivatives 
on the lipid faces, $R\pm d/2$. The two jump
conditions furnish the two parameters 
in the solution to Laplaces equation in the
bilayer, $\varphi(a<r<b) = C\ln r+ D$. 
Integrating from $r_{max}-b \gg \kappa_{o}$, 
we introduce a parameter $t$ such that
$\varphi(r_{max}) = tK_{0}(\kappa_{o}r_{max})$ 
and $\varphi^{\prime}(r_{max}) = -t\kappa_{o}K_{1}
(\kappa_{o}r_{max})$ consistent with the 
linearized solution. The parameter $t$ is then 
tuned until the solution at $r=0$ obeys 
$\varphi^{\prime}(0) \rightarrow 0$.
Solutions for $\varphi(r)$ are then used in 
the nonlinear free energy Eq. \ref{G0}.

\begin{figure}[htb]
\begin{center}
\leavevmode
\epsfysize=2.9in
\epsfbox{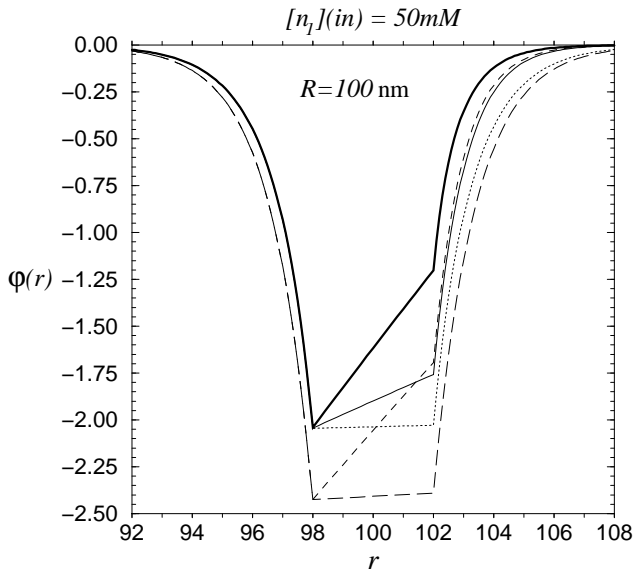}
\end{center}
\caption{Electrostatic potential 
$\varphi(r)\,(k_{B}T/e)$ near 
negatively charged interfaces
($\sigma_{a}=\sigma_{b}=-0.2\vert
e\vert \mbox{/nm}^{2}$) centered at $R=100$nm with 
50mM monovalent salt solution inside.
The dotted and thin/thick solid curves are
nonlinear Poisson-Boltzmann theory for an exterior
solution of 50mM monovalent and 
33.3mM divalent anions/cations, respectively. 
The long(short) dashed curves correspond to linear
theory  with 50mM monovalent(33.3mM divalent) ions
comprising the exterior solution. 
The linearised theory in the exterior
region does not distinguish between cations ($z_{+}=2$) 
and anions ($z_{-}=2$). }
\label{POTENTIAL}
\end{figure}

Figure \ref{POTENTIAL} plots the electrostatic potential
$\varphi(r)$ calculated from both Eq. \ref{DIFFEQ}
and its linearized
form Eq. \ref{DIFFEQLIN}.
For large radii ($R\gg \kappa^{-1}$), 
both the linear and nonlinear potentials 
$\varphi(r)$ about the interfaces at $R\pm d/2$
are essentially those of a one-dimensional theory 
and are insensitive to $R$. However, we will show that 
despite the similar behavior of $\varphi$, 
the behavior of $g_{e\ell}(R)$ for the linear and
nonlinear cases are drastically different. 

The nonlinear free energies are calculated 
for various $R$ and fitted
for large $R$ to polynomials in $R^{-1}$. The $1/R$
coefficients agree with those obtained by the analytic
method of the previous subsection. The $1/R^{2}$ 
coefficients are found to 
be drastically different from that expected from 
linear theory for all physiologically reasonable 
conditions as previously found (see Figure 1 of
Winterhalter and Helfrich 1992). 

\vspace{8mm}
\noindent {\bf Net Charge/Finite Volume Effects}

\noindent Our calculations have assumed that the solutions on both
sides of the bilayer are in chemical contact 
with large reservoirs. Applied to a closed vesicle, the
interior potential is determined by the boundary
conditions at the bilayer and the zero reference
potential at infinity (outside). Using Gauss's law on
a surface inside the bilayer ($a<r<b$) we
infer from the electric field calculated in the linear
approximation a net interior charge (all solution charges
plus $ \sigma_{a}$) per area of 

\begin{equation}
{\epsilon_{\ell}\over \epsilon_{w}}\left({\sigma_{a}\over
\kappa_{i}d}-{\sigma_{b}\over \kappa_{o}d}\right)\left[
1+O(\epsilon_{\ell}/\epsilon_{w}, 1/R)\right]
\end{equation}

\noindent for the vesicle. Even though on a per volume 
basis this is very small, the consequences of imposing
any other value inside, with the same order of magnitude,
are dramatic. These concerns are not 
simply academic for once a vesicle
is closed the interior charge may be fixed, so altering 
even the exterior electrolyte will lead to an imbalance
between the charge present and that required under the
open reservoir assumption. 

For a symmetric closed vesicle, the interior potential to
within an additive constant can be found uniquely from
the net charge in the solution; Gauss's law on the 
aqueous side of the bilayer ($r=a_{-}$) gives the
necessary boundary condition. To generalize Eq. \ref{G0} we
have only to use the expression given by Dresner (1963)

\begin{equation}
U\left[\varphi\right]= {4\pi k_{B}T\over \epsilon_{w}} 
\sum_{\alpha}N_{\alpha}\,
\ell n\, V^{-1}\int_{0}^{a}e^{-q_{\alpha}\beta e
\varphi(r)}d^{3}r
\label{USC}
\end{equation}

\noindent where $q_{\pm} = \pm z_{\pm}$. 
By demanding stationarity under variations in
$\varphi$ we find

\begin{equation}
\nabla^{2} \varphi(r) = {-4 \pi e\over 
\epsilon_{w}}\left[{n_{+}e^{e\beta\varphi(r)}
\over \langle e^{e\beta\varphi}\rangle } -
{n_{-} e^{-e\beta \varphi(r)} \over \langle e^{-e\beta
\varphi}\rangle}\right]
\label{DIFFEQSC}
\end{equation}

\noindent where we have specialized to 
monovalent ions only.
$N_{\pm}$ is the total number of
cations/anions in the interior solution, $n_{\pm} =
N_{\pm}/V$ (with $V$ the interior volume),
$\langle \ldots\rangle$ denotes $V^{-1}\int\ldots d^{3}r$.
Of course Eq. \ref{DIFFEQSC} implies Gauss's law and is
invariant under constant shifts in $\varphi$. Expanding
the potential $\varphi(r) = \delta\varphi + \langle
\varphi\rangle$ about its interior average 
value $\langle\varphi\rangle$, determined largely by the 
exterior solution ($\varphi(r\rightarrow\infty) = 0$)
and the boundary conditions, we obtain
$\nabla^{2}\tilde{\varphi}(r)
=\kappa_{i}^{2}\tilde{\varphi}(r)$,

\noindent where

\begin{equation}
\tilde{\varphi}(r) = \delta\varphi-
{k_{B}T \over e}{n_{+}-n_{-} \over n_{+}+n_{-}}.
\end{equation}


\noindent The solution to $\delta\varphi(r)$ are 
similar to the potentials derived from  Eq. \ref{DIFFEQLIN}
except for multiplicative factors and constant shifts, 

\begin{equation}
\delta\varphi(r<a) = -{2\pi a\over \epsilon_{w}
\kappa_{i}}{(n_{+}-n_{-})e \over I_{1}
(\kappa_{i}a)}I_{0}(\kappa_{i}r)
+ {k_{B}T\over e}{n_{+}-n_{-}\over n_{+}+n_{-}},
\end{equation}


\noindent from which we see $\langle\delta\varphi\rangle
= 0$. 

For purposes of illustration we evaluate $G_{e\ell}$ only
in the case $n_{+}=n_{-}$ so the interior potential is
constant and $\delta\varphi = 0$,
a feature that persists in a nonlinear
treatment. Here, $\langle\varphi\rangle \simeq
4\pi d\sigma_{a}/\epsilon_{\ell} +
O(\epsilon_{\ell}/\epsilon_{w}\kappa d)$ and is a
factor of $\sim 40$ larger than $\varphi(S)$
calculated in the previous sections; however, 
in the linear limit with $n_{+}=n_{-}$, 
the free energy given by Eq. \ref{GLIN} still holds. 
The coefficients in
a $1/R$ expansion are

\begin{equation}
\begin{array}{r}
\displaystyle C_{1} = -{2\pi d^2 \over 
\epsilon_{\ell}}\bigg[\sigma_{a}^2+{\epsilon_{\ell}\over
\epsilon_{w}\kappa_{o}d}\bigg({(\sigma_{a}+
\sigma_{b})^2\over 2\kappa_{o}d}+{3 \over
2}\sigma_{a}^2+ \\[13pt]
\displaystyle \sigma_{a}\sigma_{b}-{\sigma_{b}^{2}\over
2}\bigg)\bigg] \quad\quad
\end{array}
\end{equation}

\noindent and

\begin{equation}
\begin{array}{r}
\displaystyle C_{2}= {2\pi d^3 \over 3\epsilon_{\ell}}
\bigg[\sigma_{a}^2+{\epsilon_{\ell}\over
\epsilon_{w}(\kappa_{o}d)^2}\bigg({9
(\sigma_{a}+\sigma_{b})^2 \over 8\kappa_{o}d}+
3\sigma_{a}^2(1+\kappa_{o}d)+  \\[13pt]
\displaystyle 3\sigma_{a}
\sigma_{b}\bigg)\bigg]. \quad
\end{array}
\end{equation}

\noindent Both $C_{1}$ and $C_{2}$ are a factor of 
$\sim \epsilon_{w}/\epsilon_{\ell}$ larger than in the
conventional theory. For small $\epsilon_{\ell}$, the dominant
term arises from $\sigma_{a}\varphi(a)$ and its sign
merely says that a charged shell prefers
energetically to elongate. 

\vspace{8mm}

\noindent {\bf RESULTS}

\noindent In this section, we investigate the effects of
charge asymmetry,
ionic strength and multivalency
of the inner and outer solutions on the 
linear and nonlinear electrostatic energies. 
A free energy of the form (\ref{G0}) holds for
each side of the impermeable membrane. Energies 
and charge densities will be expressed in units of 
$k_{B}T$/nm$^{2}$ and $\vert e\vert$/nm$^{2}$ 
respectively.

\begin{figure}[htb]
\begin{center}
\leavevmode
\epsfysize=5.8in
\epsfbox{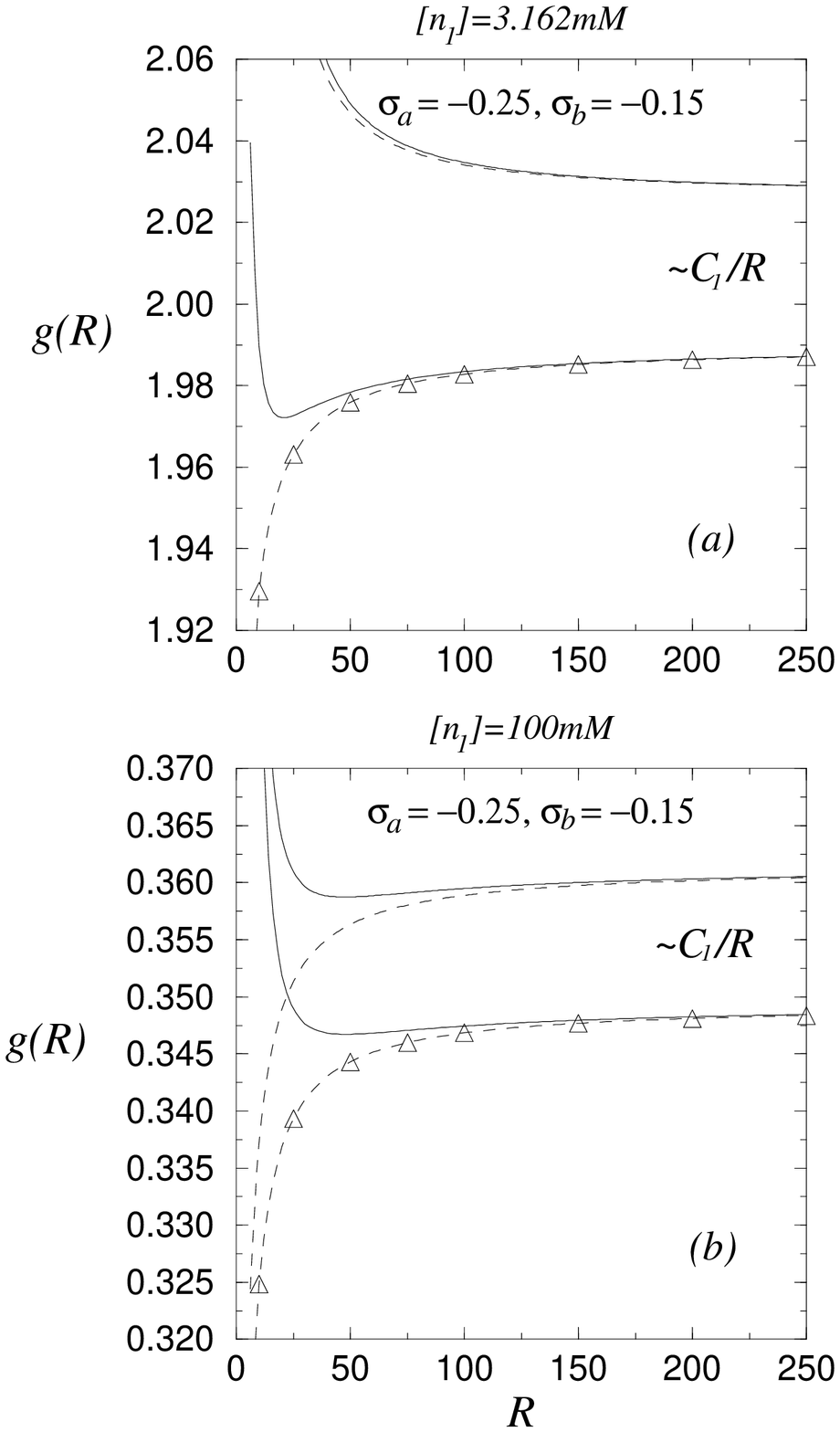}
\end{center}
\caption{Linear and nonlinear free energies per area
as a function of midplane radius
for surface charges $\sigma_{b}=-0.15$, and
$\sigma_{a}=-0.25$. Dashed lines represent
electrostatic contributions only, solid lines
depict $g_{TOT}(R)=g_{e\ell} + 6k_{B}T/R^{2}$. 
Linear solutions are shown in the upper pairs of 
curves while nonlinear Poisson-Boltzmann 
solutions are shown in the lower pairs of
curves (nonlinear $g_{e\ell}(R)$ are fits to 
the open triangles). 
(a). $10^{-2.5}\mbox{M} \simeq 3.16$mM monovalent salt 
inside and outside the vesicle. For plotting
purposes, the nonlinear energies have 
been shifted upwards by $+0.8k_{B}T/\mbox{nm}^{2}$. 
(b). $n_{1}=100$mM on both sides. 
The nonlinear energies have been
shifted by $+0.01k_{B}T/\mbox{nm}^{2}$. }
\label{GR}
\end{figure}

The form of $g_{e\ell}(R)$ is displayed in Figs. \ref{GR}
(a) and (b). Here we have for simplicity only considered
the charge per physical leaflet area ensemble with 
$\sigma_{a} = -0.25,\, \sigma_{b}=-0.15$, and monovalent
ions both interior and exterior to the bilayer.
In Fig. \ref{GR}(a), ($n_{1}=10^{-2.5}\simeq 3.16$mM),
the linear free energy $g_{e\ell}(R) = 
{1 \over 2}\int \sigma(S)\varphi(S) dS$ (dashed
line) is monotonically decreasing with increasing
$R$, whereas the nonlinear $g_{e\ell}(R)$ 
has $C_{1} < 0$. When ion concentration is increased as 
in Fig \ref{GR}(b), ($n_{1}=100$mM), 
nonlinear and linear results become
similar in that $C_{1}<0$.  In fact, for the
parameters used, the nonlinear $g_{e\ell}(R)$ shows no 
minimum in $R$ implying that in the absence of other
forces, the lipid tubule will collapse until
the ``hard wall'' limit $R\simeq d$ is approached
and the inner surface charge becomes a line. 

However, in addition to ionic forces,
other shorter-ranged electrostatic and 
entropic interactions lead to mechanical bending 
rigidities $k_m$. The solid lines in both graphs plot 
$g_{e\ell}(R) + {1\over 2}k_{m}/R^{2}$, with 
$k_{m} = 12 k_{B}T$. Under realistic physical
conditions, the electrostatic contributions to the
$1/R^{2}$ terms in the free energy, $C_{2}>0$, are 
small compared to measured mechanical $k_{m}$ values,
which in uncharged vesicles fall in the range $k_{m} \sim
2-30k_{B}T$ (Song and Waugh, 1990; Andelman 1995).
Therefore, the total free energy, for large $R$ behaves as 
 
\begin{equation}
g_{TOT}(R) \simeq C_{0}+ {C_{1} \over R}
+ {(C_{2} + k_{m}/2)\over R^{2}} + O\left(1/R^{3}\right),
\end{equation}

\noindent we will henceforth consider the $1/R^{2}$
coefficient, $C_{2}+k_{m}/2 \simeq k_{m}/2$ as an
independently measured stiffness, of mostly nonionic
contributions, and use an intermediate approximation,
$k_{m} \simeq 12k_{B}T$, where required.
The balancing of the $1/R$ and $1/R^{2}$ 
contributions controls the size scales in 
bilayer bending and membrane structures, such that 
the free energy minimizing radius is

\begin{equation}
R^{*} \simeq {k_{m} + 2C_{2} \over \vert C_{1}\vert},
\label{Rmin}
\end{equation}

\noindent for $C_{1}<0$, and the gain in free energy
(relative to the flat state) at this radius is

\begin{equation}
g_{TOT}(\infty)-g_{TOT}(R^{*}) \simeq
{C_{1}^{2} \over 2(k_{m} + 2C_{2})}.
\label{DELTAG}
\end{equation}  
  
\begin{figure}[hb]
\begin{center}
\leavevmode
\epsfysize=6.7in
\epsfbox{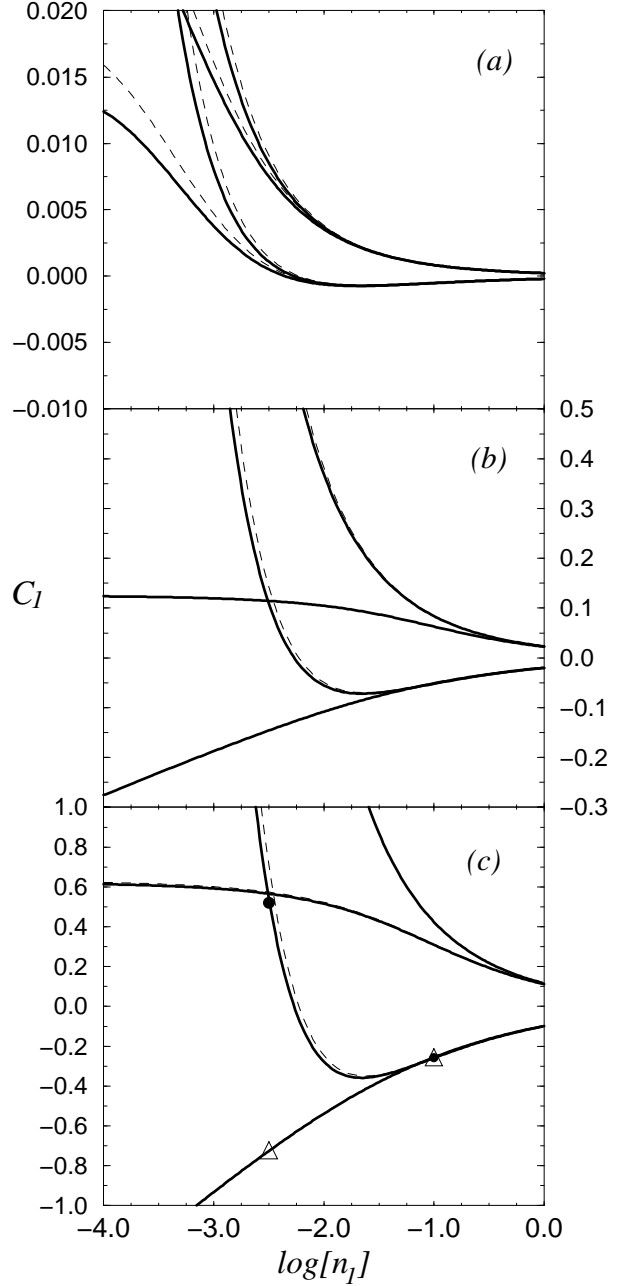}
\end{center}
\caption{$C_{1}\,(k_{B}T/\mbox{nm})$ as a function of 
monovalent ion concentration (log$_{10}$ scale,
equal both sides). (a). $\sigma_{a} =
-0.021,\,\sigma_{b}=-0.019.$ 
(b). $\sigma_{a}=-0.21,\, \sigma_{b}=-0.19.$
(c). $\sigma_{a}=-0.25,\, \sigma_{b}=-0.15.$
The filled circles and triangles indicate the
approximations and parameters used in plotting
$g_{e\ell}(R)$ in Figure \ref{GR}. Dashed(solid) curves 
show $C_{1}$ under decoupled(coupled) approximations.
The upper set of curves in each graph ($C_{1} > 0$ for 
large $n_{1}$) correspond to surface charge per midplane
area, while the lower branches ($C_{1} < 0$ for 
large $n_{1}$) correspond to surface charge conserved
with respect to area physically occupied by charges.}
\label{C1conc}
\end{figure}

In Figs. \ref{C1conc}(a), (b), and (c), we plot
the dependence of $C_{1}$ on $n_{1}$ for three different
groups of surface charges;
($\sigma_{a}=-0.021,\,\sigma_{b}=-0.019$),
($\sigma_{a}=-0.21,\,\sigma_{b}=-0.19$), and
($\sigma_{a}=-0.25,\,\sigma_{b}=-0.15$), and 
both surface charge ensembles. 

\noindent In each figure, 
linear theory is represented by curves which 
sharply increase for low $n_{1}$. The behavior of 
$C_{1}$ calculated from nonlinear theory is drastically
different for low ion concentrations, but approach as the
linear theory at high concentrations. 
The pair of curves ($C_{1}>0$ at large $n_{1}$)
corresponding to conserved charge per
midplane area remain positive. 
This is most easily seen in linear theory
from Eq. \ref{CM1} and represents the dominant
effect of increasing total interior surface charge. 
In the charge per midplane ensemble, the inner charge
$\sigma_{a}$ increases against a deeper potential
$\varphi(a)$ than $\sigma_{b}$ decreases; this electrostatic
work maintains $C_{1}>0$, thus biasing a negative radius of 
curvature (invaginations). Nonlinearities, as
expected, mitigate these effects.  Also shown is 
$C_{1}$ in the decoupled limit
($\epsilon_{\ell}/(\epsilon_{w}\kappa d) \rightarrow
0$), with dashed curves. We have checked that for
all physically reasonable parameters, $g_{e\ell}(R)$
varies by at most only a few percent when coupling 
between the two interfaces (at $R-d/2$ and $R+d/2$)
is neglected. 

The small effect of membrane coupling 
(for $\epsilon_{\ell} = 2,\, \epsilon_{w} = 80$)
is clearly demonstrated in Fig. \ref{C1conc}
down to $n_{1} \simeq 0.1$mM, especially 
at higher surface charges (compare Fig. \ref{C1conc}(a)
with Figs. \ref{C1conc}(b) and (c)) because 
the difference in surface potentials, 
$\varphi(b)-\varphi(a)$, remains
relatively small as the surface potential 
increase in magnitude nonlinearly. Since the
coupling induced contribution of the bilayer 
enters through $\varphi(b)-\varphi(a)$, its
relative importance diminishes as
$\vert\sigma\vert$ increases, particularly in
nonlinear theory (see Appendix B, Eq. \ref{BC1}). 
Also, note that since the
interior and exterior solutions are identical, a symmetry
exists, {\it i.e.} the interchange 
$\sigma_{a} \leftrightarrow \sigma_{b}$ leads to $C_{1}
\rightarrow -C_{1}$. The open
triangles in Fig \ref{C1conc}(c) indicate the parameters
used in generating $g_{e\ell}(R)$ in Fig. \ref{GR}.

\begin{figure}[htb]
\begin{center}
\leavevmode
\epsfxsize=3.4in
\epsfbox{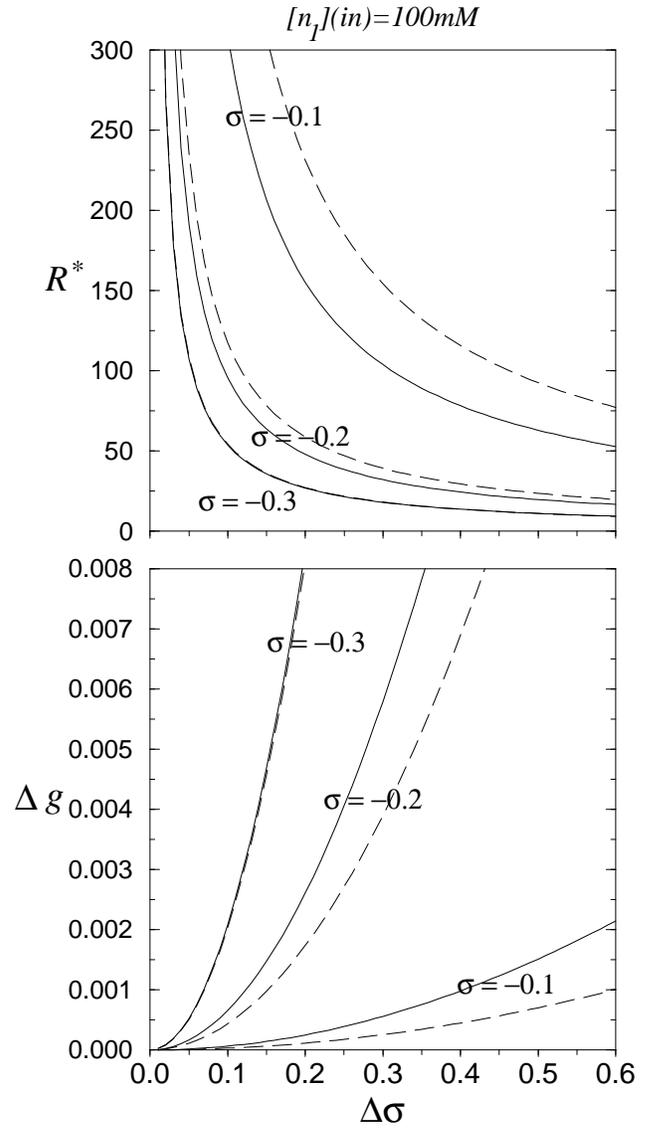}
\end{center}
\caption{ (a). Minimum energy radii 
$R^{*}\,(\mbox{nm})$  (with the assumption $k_{m} = 12k_{B}T$)
plotted as a function of charge 
asymmetry $\Delta \sigma = (\sigma_{b}-\sigma_{a})/
(\sigma_{a}+\sigma_{b})$
for mean charge $\sigma = (\sigma_{a}+\sigma_{b})/2 = -0.1, \,-0.2$, 
and $-0.3 \vert e\vert$/nm$^{2}$.
Solid(dashed) curves represent fixed charge
per leaftlet(midplane) areas. For fixed charge per
midplane area, $R^{*} > 0$ corresponds to
charge asymmetries opposite of those plotted, {\it i.e.,}
$\Delta\sigma=(\sigma_{a}-\sigma_{b})/(\sigma_{a}+
\sigma_{b})$. 
(b). The associated free energy
changes  $\Delta g(R^{*}) (k_{B}T/nm^{2})$ 
(Eq. \ref{DELTAG}).
Monovalent salt concentration is fixed
at 100mM. }
\label{RMIN}
\end{figure}

Figure \ref{RMIN} explicitly shows $R^{*}$ and
$g_{TOT}(\infty)-g_{TOT}(R^{*})$ based on Eqns.
\ref{Rmin} and \ref{DELTAG} with $k_{m} = 12k_{B}T$.
Thus, a $C_{1}$ of order
$0.1 k_{B}T$/nm is required to electrostatically induce 
radii of curvatures in the  
$\sim 50$nm range. For example, in the nonlinear cases
plotted in Figs \ref{C1conc}(b) and (c), 
at a 50mM monovalent salt concentration and conserved
charge per leaflet area, Eq.
\ref{Rmin} yields minimum free energy 
tube radii of 182nm and 37nm
respectively. The depths of these energy minima 
are given by Eq. \ref{DELTAG} as 
0.2$\times 10^{-4}k_{B}T$/nm$^{2}$ and 4.4
$\times 10^{-3}k_{B}T$/nm$^{2}$ respectively. 
Similar values are
obtained when charge per midplane area and 
the opposite charge asymmetries are considered.
Coincidentally, the magnitudes of $C_{1}$ are nearly
equal for the two conserved charge ensembles 
when $\sigma = (\sigma_{a}+\sigma_{b})/2 = -0.3$; their
magnitudes differ again as $\sigma > -0.3$.
Figure \ref{RMIN} shows that protuberances with 
radii relevant to biological systems can occur
under appropriate conditions attainable experimentally  
{\it in vitro} and {\it in vivo}.

\begin{figure}[htb]
\begin{center}
\leavevmode
\epsfxsize=3.4in
\epsfbox{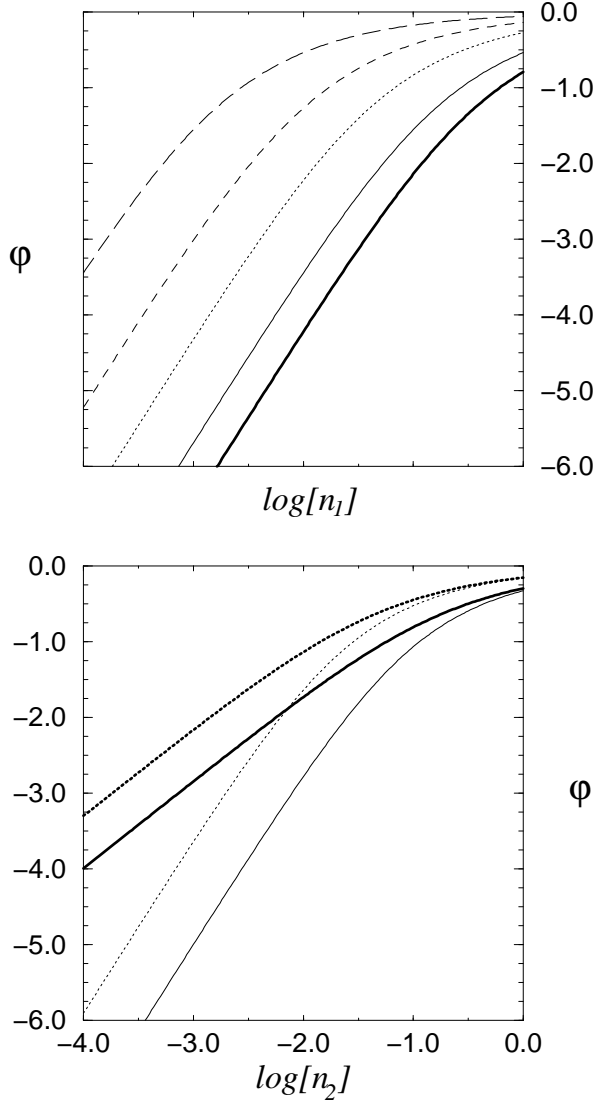}
\end{center}
\caption{Surface electrostatic potential 
at a negatively charged plane interface. 
(a). $\varphi(S)$ as a function of monovalent salt
concentration. Long dashed,
dashed, dotted, solid, and thick solid lines 
denote surface charges of $-0.02, \,-0.05,\, -0.1,\,
-0.2,$ and $-0.3\vert e \vert \mbox{/nm}^{2}$
respectively. (b). Surface potential in the presence of
a divalent salt solution. Thick(thin) lines  indicate
divalent cation(anion) solutions, and dashed(solid)
lines correspond to $\sigma = -0.1$ ($\sigma = -0.2$).}
\label{SURFPOT}
\end{figure}

The surface values, $\varphi(S)$, where
$\vert\varphi(r)\vert$ is maximal, are plotted in
Figs. \ref{SURFPOT}(a) and (b) as functions of $n_{1}$ and
$n_{2}$, respectively in order to 
assess the validity of linear theory. 
Linear theory is expected to be
accurate when $ze\beta \varphi(r) \lesssim 1$,
although the linear $C_{1}$ 
calculated in the charge per leaflet area 
ensemble seems to more accurate over a larger range of
ionic strength than that calculated in the charge 
per midplane area ensemble (see Fig. \ref{C1conc}).
The effects of divalency on $\varphi(S)$ are revealed in
Fig \ref{SURFPOT}(b). Effects of divalent
anions are quantitatively similar to those
of monovalent salts, especially if we compare at the
same concentration of monovalent cations which 
do most of the screening. However, for divalent
cations, $\vert \varphi(S)\vert$ is substantially 
reduced and linear theory is
valid over a wider range of concentrations. 
The $z_{+}=2$ species is more effective
at screening because they will balance more 
negative surface charge $e\beta\varphi(S)$ 
for the same penalty in entropy of mixing incurred.

\begin{figure}[htb]
\begin{center}
\leavevmode
\epsfxsize=3.4in
\epsfbox{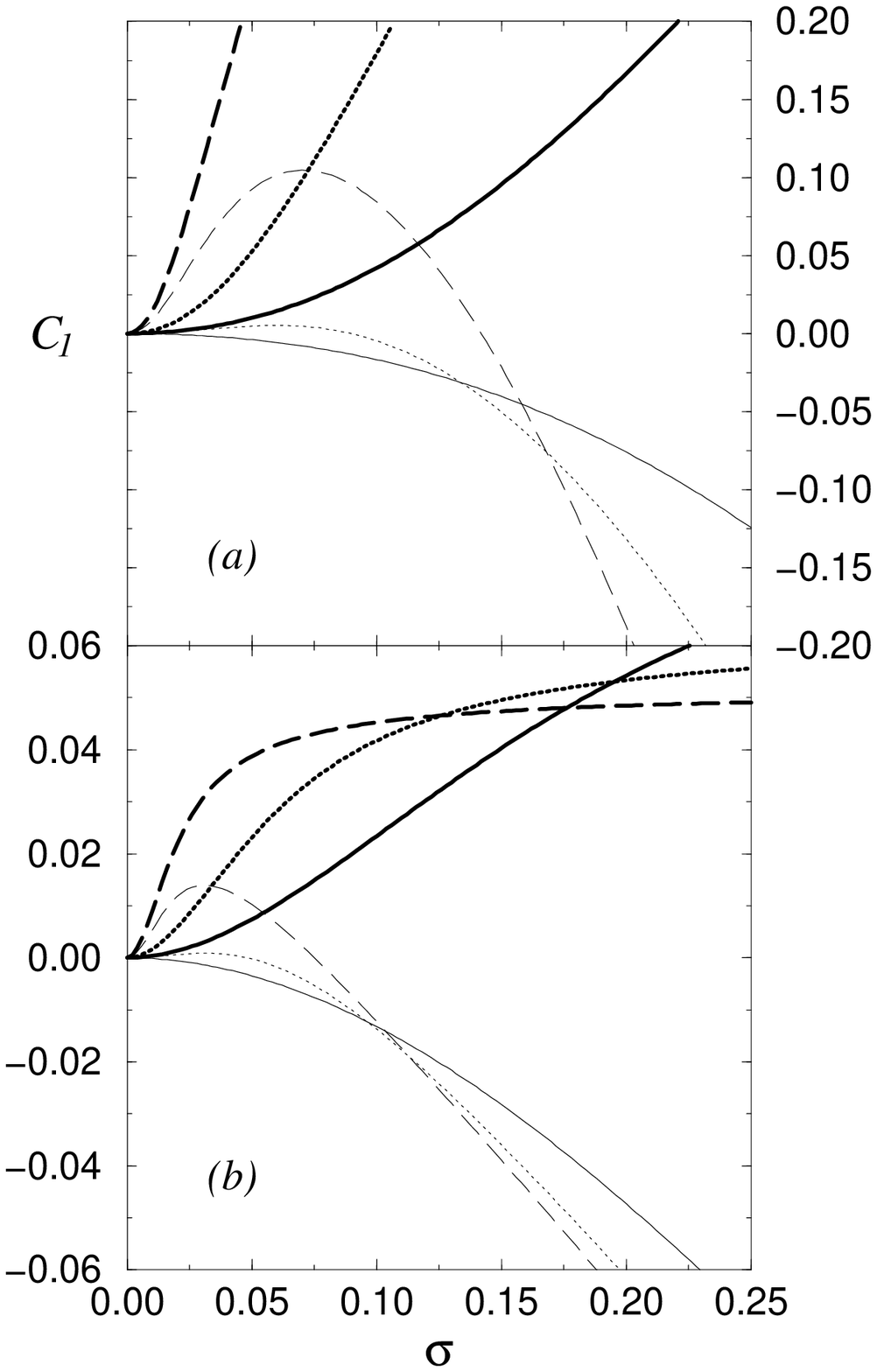}
\end{center}
\caption{$C_{1}\,(k_{B}T/nm)$ as a function of
$\sigma = -\sigma_{a}=-\sigma_{b}$ 
for various solution asymmetries. Monovalent
salt concentration in the interior is maintained  
at 1, 10, and 100mM (dashed, dotted, and solid lines
respectively) while an equiosmolar concentration 
(0.66, 6.66, and 66.6mM) of divalents make up the
exterior solution. Thick lines represent 
conserved charge per midplane area ensemble.
(a). Exterior divalent cations,
$z_{+}=2$, ({\it e.g.} CaCl$_{2}$). (b). Exterior divalent
anions, $z_{-}=2$, ({\it e.g.}Na$_{2}$ SO$_{4}$).}
\label{C1s}
\end{figure}

Differences in the inner and outer buffer 
solutions can also affect membrane bending. 
By simply changing the relative ionic strengths of the 
solutions, one can induce different screening 
and hence bilayer bending. For simplicity, we consider
the extreme case of pure monovalent and pure divalent
salt solutions. Bilayer bending due to solution 
asymmetry is demonstrated in Figures 
\ref{C1s}(a) and (b), where $C_{1}$ is shown 
as a function of $\sigma =-\sigma_{a}=
-\sigma_{b}$ for both surface charge
ensembles (charge/leaflet area, thin lines;
charge/midplane area, thick lines) 
for interior monovalent ion concentrations of 
1, 10, and 100mM  from top to bottom within each
triplet of curves. The necessary concentration of 
multivalent ions to achieve flaccid
vesicles is assumed in the exterior solution, 
{\it e.g.} $2n_{1}^{(out)} =
(z_{\pm}+1)n^{(in)}(z_{\pm})$. Here, for divalents, 
$n^{(in)}(2) = 2n_{1}^{(out)}/3$ = .666, 
6.66, and 66.66mM respectively. Figure \ref{C1s} shows that 
divalent cations (a), are more effective at 
inducing larger curvatures (larger $C_{1}$)
than divalent anions (b). 

For small surface charges in the charge per leaflet area 
ensemble, linear theory, (Eqns. \ref{KAPPA}
and \ref{C1}) is expected to hold and yields
positive(negative) slope for $C_{1}(\sigma)$ 
when $n_{1}$ is less(greater) than
$(3/2+\sqrt{2})(\ell/d)^{2}\mbox{M} \simeq 17.3$mM 
However, at larger surface charges the 
behavior of $C_{1}$ crosses
over to a negative slope for this ensemble, 
especially for higher salt
concentrations (see Fig. \ref{C1s}). Here, the
nonlinear screening in the membrane exterior 
enhances the decrease in $\vert \varphi(b)\vert$ 
relative to that of $\vert \varphi(a)\vert$ as 
$\sigma$ is increased, thus increasing the 
electrostatic energy of the interior charge layer 
relative to that of the exterior, resulting in 
bending with $C_{1}<0$. 

\begin{figure}[htb]
\begin{center}
\leavevmode
\epsfysize=3.3in
\epsfbox{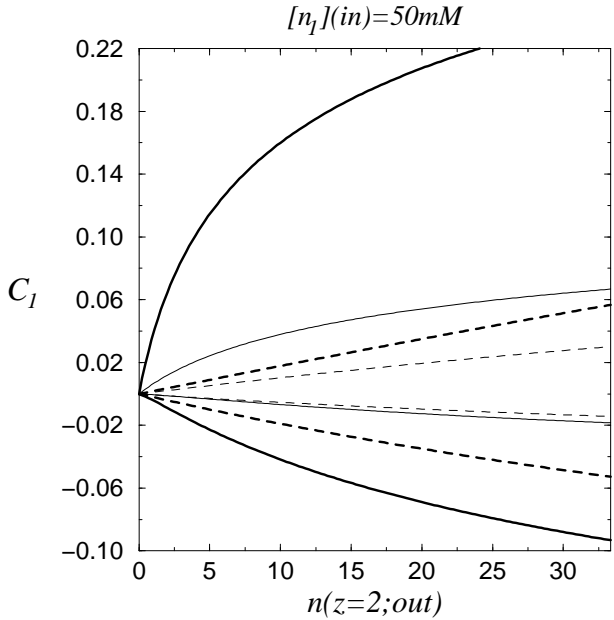}
\end{center}
\caption{$C_{1}\,(k_{B}T/nm)$ as a function of $n(z_{\pm}=2)$ 
in the exterior showing how substitution of
divalent for monovalent salt induces curvature. 
The interior solution is fixed at 50mM monovalent salt
and the solutions are equiosmolar. Thin(thick) lines
correspond to $\sigma_{a}=\sigma_{b}=
\sigma=-0.1 (-0.2)$.
The four lower curves 
($C_{1}<0$) correspond to fixed charge per leaflet area
with solid(dashed) lines corresponding to divalent 
cations(anions). The four upper curves 
correspond to fixed
charge per midplane area with other 
conventions unchanged.}
\label{C1N2}
\end{figure}

The values of $C_{1}$ corresponding 
to conserved charge per midplane area (thick curves)
increase quadratically with $\sigma$ agreeing with linear
theory, until $\sigma$ becomes large enough that 
nonlinear effects become important and saturates $C_{1}$.
The nonlinear effects are more prevalent for 
anions (Fig. \ref{C1s}(b)) for the reasons discussed
in relation to Fig. \ref{SURFPOT}.

Thus, we see that tuning the relative screening lengths 
is an effective way of inducing membrane curvature. 
Figure \ref{C1N2} shows $C_{1}$ for surfaces 
of $\sigma_{a}=\sigma_{b} = -0.1
\mbox{(thin lines) and} -0.2\mbox{(thick lines)}$ 
as the screening in the exterior solution is 
scanned. Here, the vesicle interior is
held at $50$mM monovalent salt, 
while the exterior has a varying proportion of 
monovalent and divalent salt keeping the 
vesicle flaccid 
(this requires $n_{1}^{(out)} = 0.050-3n^{(out)}(2)/2$).
The various divalent mixtures required to induce 
positive curvature ($C_{1}<0$, tube growth), 
or invaginations ($C_{1}>0$) and the sensitivity of 
$C_{1}$ to divalent concentrations are clearly shown.
Since $\sigma_{a}=\sigma_{b}$, exchange of the solutions 
on the two sides of the membrane interchanges
$C_{1} \rightarrow -C_{1}$.  

\begin{figure}[htb]
\begin{center}
\leavevmode
\epsfxsize=3.4in
\epsfbox{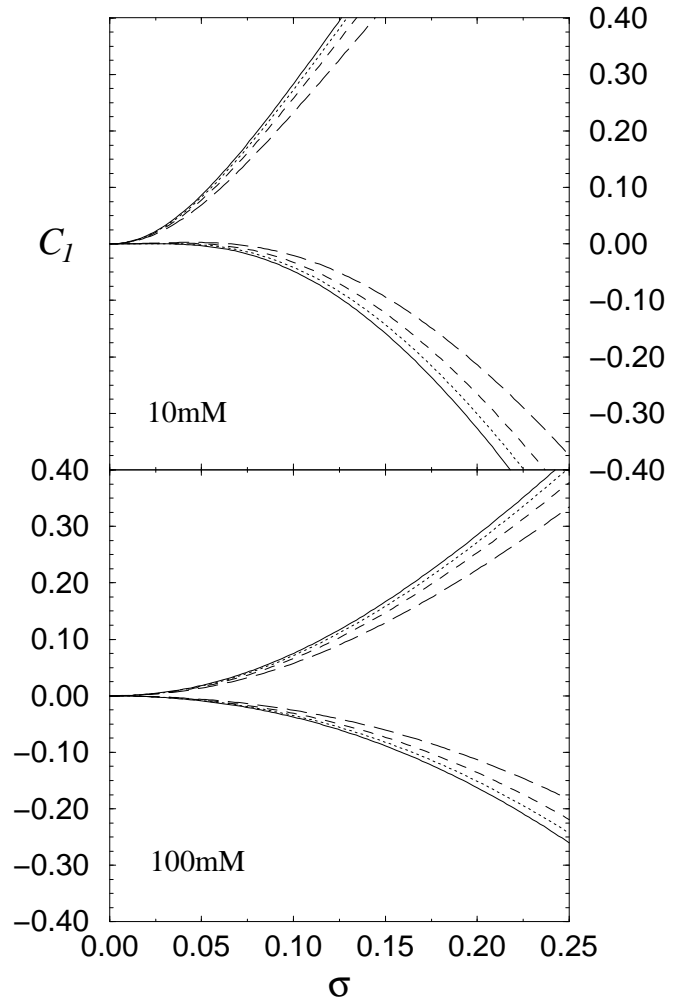}
\end{center}
\caption{$C_{1}\,(k_{B}T/nm)$ as a function of $\sigma=-\sigma_{a}
=-\sigma_{b}$ for 
exterior cationic multivalencies, $z_{+} = $3(log dashed), 
4(dashed), 5(dotted), and 6(solid) lines.
(a). $n_{1}^{(out)} = 10$mM. (b). $n_{1}^{(out)}
 = 100$mM. }
\label{MULTI}
\end{figure}

Finally, in Fig. \ref{MULTI} we plot $C_{1}$ as a
function of $\sigma = -\sigma_{a}= -\sigma_{b}$ for 
vesicles with interior monovalent salt 
exterior higher multivalent ions ($z_{\pm} > 2$).
The influence of higher valencies 
on $C_{1}$ is modest except at high 
surface charges or low total ionic strengths.

\vspace{8mm}

\noindent {\bf DISCUSSIONS AND CONCLUSIONS}

\noindent In this paper we have presented calculations which
suggest that electrostatic forces can control lipid membrane
bending under realistic experimental and
physiological conditions. Membrane deformations 
can be induced by an aqueous solution asymmetry 
between vesicle interior and exterior as well as by
charge asymmetry between the two bilayer leaflets. 
Radii of curvature of the membrane
bending can be in the neighborhood of
50-100nm typically seen in biological 
processes such as vesicle budding, 
tubulation of ER and Golgi
bodies, and endo/exocytosis. 

Tubulation and growth of necks from vesicles requires 
the membrane to nucleate such protuberances.
Electrostatic changes in free energy alone 
can yield qualitatively reasonable conditions 
for the formation of tube-like
structures from a flat bilayer membrane. 
For example, if $\Delta g_{TOT}(R^{*} \simeq
40\mbox{nm}) \simeq 0.0005 k_{B}T/$nm$^{2}$,
the total free energy decrease is
roughly $0.11 k_{B}T$/nm length.
Therefore, a flat membrane is stable against 
tube fluctuations of height 
$L \lesssim 10$nm. This critical  
length will be slightly greater due to
the additional bending energy cost at the tube 
base; however, from experimental electron microscopy 
images (Mui {\it et al.}, 1996),
the base can be approximated with a portion of a torus 
of cross-sectional radius $\sim 10$nm. 
Using a bending rigidity of
$k_{m} \simeq 12k_{B}T$, and the fact that this toroidal
section has both positive and negative curvatures, 
we find that the total mechanical
bending rigidity is qualitatively small 
and does not affect the energetics
appreciably such that once a fluctuation exceeds $\sim
10$nm, it will continue to extend and lower
$g_{e\ell}$.

Magnitudes of membrane surface 
charges experimentally measured indicate
that our canonical estimate of $-0.2\vert e\vert$/nm$^{2}$
is a reasonable physiological value. Surface charges
measured in plant vesicles using particle electrophoresis
and dye fluorescence range from $-0.03\vert
e\vert/\mbox{nm}^{2}$ to $-0.24 \vert e\vert/\mbox{nm}^{2}$
(Sack {\it et al.,} 1983; Chow and Barber 1980). These are 
averaged charge densities; higher concentrations of charged
lipid could be recruited to incipient buds (phase
separation) if electrostatics are playing a role. 
The numerous detailed chemical mechanisms of lipid-solvent
interactions have not been modeled. The variation of 
lipid pK$_{a}$'s with solution ionic strengths, the
nonelectrostatic binding of cations to membrane surfaces,
and the hydrogen bonding among lipid headgroups can all
affect the effective surface charge and is discussed
by Tocanne and Teissi\'{e} (1990). 

\begin{figure}[htb]
\begin{center}
\leavevmode
\epsfysize=4.2in
\epsfbox{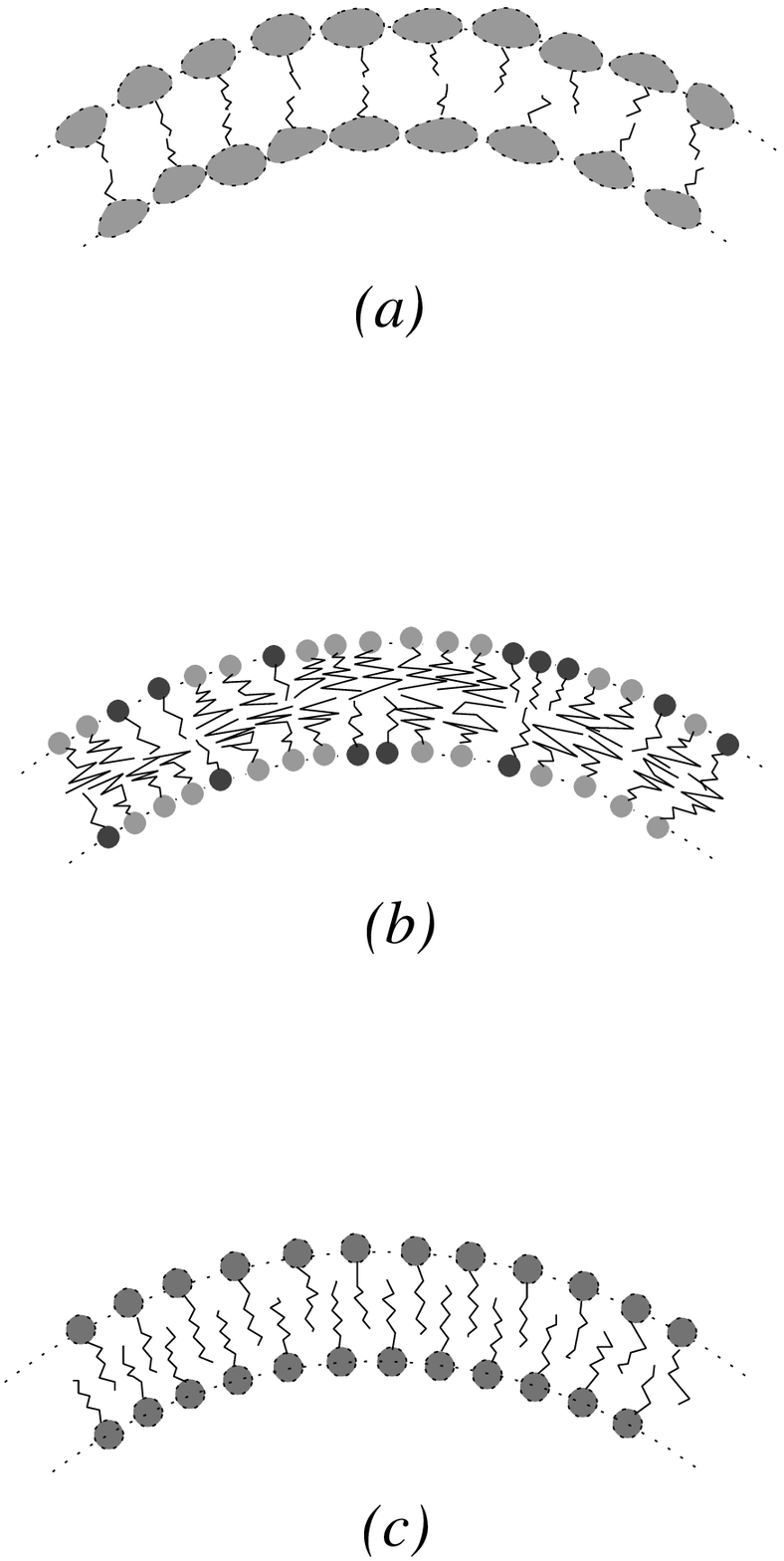}
\end{center}
\caption{Microscopic models governing the distribution of
charge and the neutral surface. 
(a). Bulky heads sterically fix charge per inner and
out leaflet areas. (b). Small heads and entropically 
interacting acyl chains distribute charges at the heads
according to midplane area. (c). Charge per midplane area
preserved owing to interdigitated acyl chains. }
\label{MICRO}
\end{figure}

The response of surface charges to bending 
is also crucial in determining the electrostatic 
contributions to the free energy. The two 
extreme cases examined correspond to
charge distributions which are lipid tail or head 
controlled and are depicted in Figs. \ref{MICRO}. 
For large head groups where lipid packing is governed by 
steric and electrostatic interactions among head groups,
the surface charge is approximately held constant
as the bilayer curvature is varied. 
Conversely, when the tails occupy a thermodynamic 
area larger than the heads, the surface charge 
is approximately fixed with respect to 
the area defined by the midplane
of the bilayer. Furthermore, other special 
phases may be important in determining charge 
distribution. For example, if the solvent contains 
glycerol or alcohols, the acyl chains of PC bilayers
become interdigitated (Gennis, 1989) 
as shown in Fig. \ref{MICRO}(c). 
If this coupling is tight, the surface charge on the
leaflets would be determined by the
midplane area. 

It is interesting to note that 
the modes of charge conservation mentioned above can
be correlated with a filling parameter   
defined by Israelachvilli (1979),

\begin{equation}
f= v/a\ell,
\end{equation}

\noindent where $v$ is the effective volume of
the lipid tail, $a$ is the thermal area
of the head or the tail group near the head, 
and $\ell$ is the vertical thermal length of the tail. 
For $f\simeq 1$, the lipids are schematically 
represented by cylinders. For $f>1$ and $f<1$, the
lipids behave microscopically approximately 
as cones and inverted cones, respectively. 
The $f>1$ (``tail packed'') or inverted cone
$f<1$ (``head packed'') structures will probably be 
associated with fixed surface charge with respect
to midplane area and fixed surface charge 
with respect to leaflet interfaces ($r=a,\,b$) 
ensembles, respectively. Depending on the 
chemical composition of the bilayer, an intermediate 
neutral surface between $R$ and $a,\,b$ is also
possible (Petrov and Bivas 1984, Gennis 1989).

Correlating the cartoon in Figs. \ref{MICRO} and the
arguments above with bilayer chemistry 
may serve as an important guide in
developing {\it in vitro} experiments and understanding
biological processes. Studies of the phases of 
concentrated lipid solutions suggest that $f$ is 
related to lipid micellular, planar, and inverted
micellular structures.  For example, lipids 
such as PC, PS, PI, PG, and Sphingomyelin
at nonacidic conditions and in the absence of 
divalent cations (Kates and Manson, 1984)
form stable planar bilayers,
implying that inter-lipid interactions 
through the tail and headgroups are comparable.
Lysophospholipids, on the other hand, form micelles with the 
headgroups pointing outward into the aqueous phase,
and can be modelled by inverted cones, with the cone
apex at the midplane of the bilayer, commensurate with
strong headgroup interactions. 

In general, factors that 
increase the effective acyl chain area relative to that of
the headgroups favors H$_{II}$(inverted hexagonal)
phase over the L$_{\alpha}$(planar) phase (Cullis {\it et
al.} 1985). As in the case of surface charge, chemical
conditions affect how lipids are packed in a bilayer. 
For example, a low pH tends to increase
headgroup association through 
hydrogen bonding, (Boggs 1984) 
and cations, such as Mg$^{+2}$ and
Ca$^{+2}$, also decrease headgroup size by 
dehydrating them. This enhances the likelihood of the  
H$_{II}$ phase. Lee, Taraschi and Janes (1993) have 
qualitatively applied the lipid shape concept to 
the POPtdEtn/PtdEth binary lipid mixture. Under their
conditions, PtdEth has a large tail area which promotes
the formation of inverted hexagonal phases in the lipid
mixture (Lee, Taraschi, and Janes, 1993). In our study, 
vesicles containing similar types of lipids may be expected
to conserve surface charge with respect to midplane 
area $2\pi RL$.

Probably the least accurately measured 
parameter in bilayer electro-mechanical models is 
the coefficient of $1/R^{2}$ in $g_{TOT}(R)$.
Measurements using various methods have yielded
scattered results. For example, in
phosphotidylcholines (PC), large values of $k_{m}$ are
obtained when employing tube bending measurements, 
intermediate values are obtained when a fluctuation
mode analysis is performed, and low values are
extracted when electric field deformation of spherical
vesicles is analyzed (Andelman 1995). For example, 
Song and Waugh (1993) measured $k_{m}\simeq 28k_{B}T$ 
for SOPC by mechanically pulling tethers. This value
increased by $\sim 3$-fold when $\sim 45\%$ 
cholesterol was added. Mode fluctuation measurements
by Mutz and Helfrich (1990) on lipid vesicles 
$\sim 10\mu m$ yielded $k_{m} \simeq 4k_{B}T, 
\, 28k_{B}T$ and $100k_{B}T$ for
galactosyldiglyceride, DMPC, and DMPC $+$ 30\%
cholesterol bilayers respectively. Finally  
electric field deformation studies by 
Duwe {\it et al.} (1990) on DGDG and Egg Yolk PC 
vesicles yielded $k_{m} \simeq 2 k_{B}T$ and $k_{m}
\simeq 5k_{B}T$ respectively. All of these measurements were
performed under neutral pH and low salt concentrations
where electrostatic contribution to the 
total bending rigidity $k_{m}+2C_{2}$ may be important.
However, without careful experimental control of the 
solution ionic strengths (and surface charges), variation in
$C_{2}$ can lead to the discrepancies reported. The enhancement of
membrane bending stiffness upon surface association of 
uncharged polymers has also been studied by
Evans (1996). Throughout this
paper we have simply subsumed all these effects
into an intermediate value of
$k_{m}+2C_{2} \simeq 12k_{B}T$, where $C_{2} \ll k_{m}$ 
at the higher salt concentrations considered.

Song and Waugh (1990) also measured bending stiffnesses
of artificial mixed POPS-SOPC vesicles as a function of
surface charge by varying the composition of charged POPS.
They found no difference in total bending stiffness 
for $0\, ,2\, ,$ and $16$\% POPS ($\sim
\mbox{35}k_{B}T$). However, the solution ionic
strengths and effective surface charges
in these experiments were not precisely controlled and 
assuming surface charges appropriate to 
the added amounts of POPS, even screening lengths $\sim
30$nm can cause $C_{2}$ to saturate at a maximal value $\ll
k_{m}$ (Winterhalter and Helfrich 1992).

Experiments on artificial liposomes hitherto have not 
carefully considered the effects of solution 
ionic strength on deformation, tubulation and budding. 
In the experiments of Mui {\it et al.} (1996), 
a pH gradient across a bilayer was used to flip lipids with 
different pKa's across from one leaflet to
the other. Tubulation was induced which the authors 
attributed to a leaflet area imbalance driven by the
lipid exchange. However, as shown by 
Hope {\it et al.} (1989), pH induced transport of 
lipids across the 
bilayer also changes the relative charges 
between the bilayer leaflets.
These experiments therefore do not isolate 
the electrostatic effects we have calculated, 
even though, as we have shown, the 
electrostatic component may play 
a significant role in vesicle shape changes. 
Using natural Golgi bodies, Cluett {\it et al.} (1993)
performed experiments with Brefeldin A (BFA) 
and approximately 50mM ion concentration. 
Tubules up to $7\mu$m in length 
grew when BFA was added to prevent binding of coat
proteins and budding. The tubes were of the same
size, $\sim 70$nm, as typical budding vesicles,
suggesting that a common controlling factor such as
electrostatics is not unreasonable. 

Besides continuum electrostatics, 
there are numerous other chemical and biological 
effects which can alter the mechanical properties
of a bilayer. In particular, ion binding and
hydrogen bonding effects have not been considered.
However, experiments hitherto have 
not carefully controlled parameters affecting
even the electrostatics: solution 
ionic strength and surface charge. 
We propose that {\it in vitro} experiments on 
large vesicles be performed under flaccid conditions 
with ionic strength as well as pH carefully measured.
Systematically varying surface charge may 
also be appropriate. 

Our calculations have shown how adding 
neutral dopants to a bilayer can indirectly
induce morphological changes of electrostatic
origin. For example, if 
enough cholesterol, a rigid molecule with a 
relatively small head group, is incorporated 
equally in the bilayer leaflets, upon bending,
phospholipid charges will be conserved with respect to
a neutral surface closer to the the midplane radius $R$
as indicated by Figs. \ref{MICRO}b, c.
Although $k_{m} >0$ is also expected to change 
in magnitude, the sign of $C_{1}$ is 
very sensitive to how charge is conserved. 
With asymmetry in $\sigma_{a,b}$ (Fig.
\ref{C1conc}) or interior/exterior screening
(Fig. \ref{C1N2}), the addition of 
cholesterol can change the 
sign of $C_{1}$ and determine whether tubules 
grow outward or invaginate. A gradient in
lipid composition occurs biologically, 
for example with cholesterol in the Golgi-ER 
membranes (Bretscher and Munro 1993). 
Such membranes are constantly
tubulating, budding, and recycling and their 
local cholesterol content may determine the 
neutral surfaces which govern the 
electrostatic component of these processes.

We have also shown that provided a pure lipid bilayer is
impermeable to ions, the conservation of charge in the
interior of a closed vesicle can completely alter the 
electrostatic energies calculated assuming thermodynamic
reservoirs. Only for a certain interior charge density
specified to $O(\sigma_{a}/R)$ will the conventional
result apply, though for a few tubes growing from a 
sphere of much larger area our charge imbalance analysis
should be applied to the sphere only. For {\it in vitro}
studies of pure bilayer vesicles it is experimentally 
difficult to control the interior charge to the required
accuracy and thus large electrostatic effects are likely
to be present at least initially before the inner and 
outer solutions have equilibrated.

For an interior charge of order $\sigma_{a,b}$ per area
the internal potential will be of order 
$(\epsilon_{w}/\epsilon_{\ell})k_{B}T/e$, which is absurdly
large for biological membranes, and even for an
artificial bilayer will cause ions to traverse it. 
If electrogenic ion pumps act to maintain a 
electronic potential
difference of several $k_{B}T$ across an organelle's
membrane there may still be an effect on the optimal
curvature comparable to what we have calculated
in the Results section. Our calculations are most apt for a
membrane where pores allow small inorganic ions to
equilibrate while multivalent proteins are localized to
one side and control the bending.

Finally, we have verified that the effects of multivalent 
species ($z_{\alpha}>2$) in the surrounding buffer solution
are similar to those of the divalent solutions 
which we have treated in more detail. 
This has implications for vesicle budding 
assisted by adaptin/clathrin or dynamin proteins, whose 
molecular charges can interact and screen those 
at the bilayer surfaces.  

\vspace{8mm}

\noindent {\bf ACKNOWLEDGEMENTS}

\noindent TC thanks J. E. Evanseck for related discussions.
TC and EDS acknowledge the support of NSF grant 
DMR-9300711. MVJ acknowledges support from 
NSF grant DMR-9215231.

\vspace{8mm}
\noindent {\bf APPENDIX A}

\noindent The solutions to the linear equation \ref{DIFFEQLIN}.
in the geometry of Fig. \ref{FIG1} under appropriate 
electrostatic boundary conditions are displayed:

\begin{equation}
\begin{array}{l}
\displaystyle \varphi(r<a) = 4\pi \sigma_{a} 
\kappa_{o}b aK_{1}(\kappa_{o}b)\bigg[{\epsilon\over
\epsilon_{w}}\bigg({\sigma_{b}\over
\sigma_{a}\kappa_{o}a} + {1 \over \kappa_{o}b}\bigg)
{K_{0}(\kappa_{o}b)\over K_{1}(\kappa_{o}b)}+ \\[13pt]
\displaystyle \quad\quad\quad\quad\hspace{4cm}
\ln \left({b\over a}\right)\bigg]{ I_{0}(\kappa_{i}r)
\over D(\kappa; a,b)} \\[15pt]
\displaystyle \varphi(a<r<b) = 4\pi \sigma_{b}
{\kappa_{i}abI_{1}(\kappa_{i}a)K_{1}(\kappa_{o}b)
\over D(\kappa; a,b)}\ln r 
+ \mbox{constant} \\[15pt]
\displaystyle \varphi(r>b) = 4\pi \sigma_{b} 
\kappa_{i}a bI_{1}(\kappa_{i}a)\bigg[{\epsilon\over
\epsilon_{w}}\left({\sigma_{a}\over
\sigma_{b}\kappa_{i}b} + {1 \over
\kappa_{i}a}\right){I_{0}(\kappa_{i}a)\over
I_{1}(\kappa_{i}a)}+ \\[13pt]
\displaystyle \quad\quad\quad\quad\hspace{4cm}
\ln \left({b\over a}\right)
\bigg]{K_{0}(\kappa_{o}r)
\over D(\kappa;a,b)},
\label{LINSOLN}
\end{array}
\end{equation}

\noindent where

\begin{equation}
\begin{array}{r}
\displaystyle D(\kappa;a,b) \equiv \epsilon_{w}
\kappa_{i}a\kappa_{o}bI_{1}(\kappa_{i}a)K_{1}
(\kappa_{o}b)\bigg[{\epsilon_{\ell}\over
\epsilon_{w}}{1\over \kappa_{o}b}{K_{0}
(\kappa_{o}b) \over
K_{1}(\kappa_{o}b)} + \\[13pt]
\displaystyle \quad\quad\quad\quad {\epsilon_{\ell}\over
\epsilon_{w}}{1 \over \kappa_{i}a}
{I_{0}(\kappa_{i}a)
\over I_{1}(\kappa_{i}a)} + \ln \left({b\over
a}\right)\bigg].
\end{array}
\end{equation}

\vspace{8mm}
\noindent  {\bf APPENDIX B}

\noindent Write the solution to Eq. \ref{DIFFEQ2} as an 
expansion in $1/R$; $\varphi = \varphi_{0} + 
\varphi_{1}$, where  $\varphi_{0}$ is the 
solution to the one dimensional Poisson-Boltzmann 
equation for a flat interface, with boundary conditions 
appropriate to surface charges $\sigma_{a,b}$
and $\varphi_{1} 
= O(1/R)$. Expanding $G_{e\ell}$
in for example the bilayer $(a<r<b)$ and outer 
regions $(r\geq b)$,

\begin{equation}
\begin{array}{l}
\displaystyle
{ G_{+}+G_{\ell}\over 2\pi L} \simeq  \sigma_{b} 
b(\varphi_{0}(b)+\varphi_{1}(b)) - 
{\epsilon_{w}\over 4\pi}\int_{b}^{\infty}
\bigg({1\over 2}(\partial_{r}\varphi_{0})^2 +\\[13pt]
\displaystyle 
\hspace{8mm}(\partial_{r}\varphi_{0})(\partial_{r}\varphi_{1}) +
U\left[\varphi_{0}\right] + \varphi_{1}
U^{\prime}\left[\varphi_{0}\right]\bigg) 
r\,dr - \\[13pt]
\displaystyle {\epsilon_{\ell} \over 4\pi}\int_{a}^{b}
\left({1\over 2}(\partial_{r}\varphi_{0})^2 +
(\partial_{r}\varphi_{0})(\partial_{r}
\varphi_{1})\right)r\,dr+O(1/R).
\end{array}
\label{APPENDB1}
\end{equation}

\noindent Using the identity 
$(\partial_{r}\varphi_{0})(\partial_{r}\varphi_{1}) = 
\nabla\cdot(\varphi_{1}\nabla\varphi_{0})-
\varphi_{1}\nabla^{2}\varphi_{0}$, and 
recalling that to the order in 
$R$ to which we are working, we can approximate 
$\nabla^{2}\varphi_{0} \simeq \partial_{r}^{2}
\varphi_{0} \equiv 
U^{\prime}\left[\varphi_{0}\right]$ in the first 
integral, one sees that the
$U^{\prime}$ terms cancel. The total derivatives in both 
integrals reduce to surface terms which cancel
$\sigma_{b}b\varphi_{1}(b)$ and the corresponding term 
at the inner surface  when $G_{-}$ is included. This is
no accident since the electrostatic boundary conditions
just express the stationarity of (\ref{G0}) with respect
to variation in the surface value of $\varphi$. In fact
since $\partial_{r}\varphi_{0}$ satisfies boundary
conditions based on $\sigma_{a,b}$, $\varphi_{1}(a,b)
\equiv 0$. Our argument does establish that
$\varphi_{0}$ could satisfy the $R \rightarrow \infty$
boundary condition for the fixed charge per midplane
area ensemble without affecting the $1/R$ coefficient.
Of course the explicit $\sigma_{a,b}$ that occurs in
$G_{\pm}$ should be correct for the particular ensemble.

To extract the $R$ dependence from the integrals over
$\varphi_{0}$ we write $rdr \rightarrow \pm(R\pm d/2
\pm z)dz$ in $G_{\pm}$ and expand.  Incidentally this
argument shows that for a sphere  the coefficient of
$1/R$ is doubled, due entirely to the variation of
surface area with radius. The integral within the
bilayer is symmetric about $R$ and does not contribute
to the $1/R$ coefficient. For numerical purposes 
it is convenient to use Eq. \ref{DIFFEQ3} and express
the remaining integrals in Eq. \ref{gTOT} using 
$\varphi_{0}$, which is a monotonic function of 
$r$, as the independent variable, {\it i.e.},

\begin{equation}
\int_{0}^{\infty}dz {\cal E}\left[\varphi_{0}\right] = 
\int_{0}^{\infty}dz (\partial_{z}\varphi_{0})^2 =
\int^{0}_{-\vert\varphi_{0}(S)\vert}
d\varphi_{0}\sqrt{2U\left[\varphi_{0}\right]},
\label{E}
\end{equation}

\noindent and similarly,

\begin{equation}
\int_{0}^{\infty}\!\!\! z\,dz {\cal E}
\left[\varphi_{0}\right] =\! 
\int_{-\vert\varphi_{0}(S)\vert}^{0}\!\!\!\!\!\!d\varphi_{0}
\left[{1 \over\sqrt{2U\left[\varphi_{0}\right]}}
\int_{-\vert\varphi_{0}\vert}^{0}\!\!\!\!\!\!d\varphi_{0}^{\prime}
\sqrt{2U\left[\varphi_{0}^{\prime}\right]}
\right].
\label{Ez}
\end{equation}

\noindent The relevant boundary values
$\varphi_{0}(S) \equiv \varphi_{0}(a),\,
\varphi_{0}(b)$, are found from the two boundary
conditions which yield,

\begin{equation}
\begin{array}{l}
\displaystyle
{\epsilon_{\ell}\over
d}\left(\varphi_{0}(b)-\varphi_{0}(a)\right)
-\epsilon_{w}\sqrt{2U\left[\varphi_{0}(b)\right]} =
4\pi\sigma_{b} \\[13pt]
\displaystyle -{\epsilon_{\ell} \over d}
\left(\varphi_{0}(b)-\varphi_{0}(a)\right)
-\epsilon_{w}\sqrt{2U\left[\varphi_{0}(a)\right]}
=4\pi\sigma_{a},
\label{BC1}
\end{array}
\end{equation}

\noindent where $\sigma_{a,b} < 0$ and the sign of the
square-root terms have been chosen accordingly.
The solution of the this simultaneous 
system of equations gives
$\varphi_{0}(a),\,\varphi_{0}(b)$ to be used in the 
limits in integrals (\ref{E}) and (\ref{Ez}).
 
If we had imposed boundary conditions that the charge 
per midplane area was conserved during bending, the only
change in Eq. \ref{APPENDB1} would be to replace
the multiplicative factor of $b$ with 
$R$ in the first term and of course to interpret
$\sigma_{a,b}$ as the charge per midplane area. 
One can continue to employ the boundary conditions
(\ref{BC1}) even though $\sigma_{a,b}$ are no longer the
correct lipid-solution surface charges, by the argument
already given that the surface value of $\varphi_{1}$
drops out. 
The only change to 
Eq. \ref{gTOT} is to remove the $d/2R$
factor multiplying $\sigma_{a,b}$.

\newpage

\noindent \hspace{-5mm}{\bf REFERENCES}

\vspace{4mm}

\noindent \hspace{-5mm} Alberts, B., Bray, D., Lewis, J.,
Molecular Biology of the Cell,
Garland Publishing NY, 1994.

\vspace{4mm}

\noindent \hspace{-5mm} Andelman, D. 1995. 
Electrostatic Properties of Membranes: 
The Poisson-Boltzmann theory. {\it In}
Handbook of Biological Physics, v. 1. Structure and
Dynamics of Membranes, Generic and Specific
Interactions. R. Lipowsky and E. Sackmann, editors.
Elsevier Science Publishers B.V.: Amsterdam. 603-642.

\vspace{4mm}
\noindent \hspace{-5mm} Ben-Tal, N., Honig, B., Peitzsch, R. M.,
Denisov, G., and McLaughlin, S. 1996. 
Binding of Small Basic Peptides to Membranes Containing 
Acidic Lipids:Theoretical Models and Experimental
Results. {\it Biophys. J.} 71: 561-575.

\vspace{4mm}
\noindent \hspace{-5mm} Bretscher, M. S., and Munro, S. 1993.
Cholesterol and the Golgi Apparatus. {\it Science} 261:
1280-1281.

\vspace{4mm}
\noindent \hspace{-5mm} de Camilli, P., Emr, S. D., 
McPherson, P. S., and Novick, P., 1996.
Phosphoinositides as regulators in membrane traffic. 
{\it Science} 271: 1533-1539.

\vspace{4mm}
\noindent \hspace{-5mm} Chow, W. S., and Barber, J. 1980.
Salt-dependent changes of 9-aminoacridine fluorescence
as a measure of charge density of membrane surfaces.
{\it J. Biochem. Biophys. Methods}  3:173-185.

\vspace{4mm}
\noindent \hspace{-5mm} Cluett, E. B., {\it et al.} 1993. 
Tubulation of Golgi membranes {\it in vivo} and {\it in 
vitro} in the absence of
Brefeldin A, {\it J. Cell Biology}. 120: 15-24.

\vspace{4mm}
\noindent \hspace{-5mm} Dresner, L. 1963. A variational
Principle for the Poisson-Boltzmann Equation.
Activity Coefficient of a Salt in a Charged
Microcapillary. {\it J. Chem. Phys.} 67: 2333-2336.

\vspace{4mm}
\noindent \hspace{-5mm} Duwe, H. P., Kaes, J., and Sackmann, 
E. 1990. Bending elastic moduli of lipid bilayers:
modulation by solutes.
{\it J. Phys. France}. 51: 945-962.

\vspace{4mm}
\noindent \hspace{-5mm} Eibl, H., and Blume, A. 1979. The influence
of charge on phosphatidic acid bilayer membranes. 
{\it Biochim et Biophys. Acta.} 553: 476-488

\vspace{4mm}
\noindent \hspace{-5mm} Evans, E. private communication.

\vspace{4mm}
\noindent \hspace{-5mm} Gennis, R. B. 1989. 
Biomembranes: Molecular 
Structure and Function. Springer-Verlag, 
New York.

\vspace{4mm}
\noindent \hspace{-5mm} Gruenberg, J. and Maxfield, F. R., 1995.
Membrane transport in the endocytic
pathway, {\it Current Opinion in Cell Biology.} 7: 
552-563.

\vspace{4mm}
\noindent \hspace{-5mm} Hope, M. J., Redelmeier, T. E.,
Wong,  K. F., Rodrigueza, W., and Cullis, P. R. 1989.
Phospholipid Asymmetry in Large Unilamellar Vesicles
Induced by Transmembrane pH Gradients. 
{\it Biochem.} 28: 4181-4187.

\vspace{4mm}
\noindent \hspace{-5mm} Hui, S. W. 1993. Lipid Molecular Shape
and High Curvature Structures. {\it Biophys. J.}
65: 1361-1362.

\vspace{4mm}
\noindent \hspace{-5mm} Israelachvili, J., Marcelja, S., and Horn, R. G.
1980. Physical principles of membrane organization. {\it
Quart. Rev. Biophysics.} 13:121-200.

\vspace{4mm}
\noindent \hspace{-5mm} Lee, Y.-C., Taraschi, T. F., and Janes, N. 1993.
Support for the Shape Concenpt of Lipid Structure Based on
Headgroup Volume Approach. {\it Biophys. J.} 65: 1429-1432.

\vspace{4mm}
\noindent \hspace{-5mm} Boggs, J. M. 1984.
Intermolecular Hydrogen Bonding Between
Membrane Lipids. {\it In} Biomembranes v.
12. Membrane Fluidity. M. Kates and L. A.
Manson, editors. Plenum Press, New York.


\vspace{4mm}
\noindent \hspace{-5mm} Lipowsky R. 1993. 
Domain-induced budding of fluid membranes
{\it Biophysical J.} 64: 1133-1138.

\vspace{4mm}
\noindent \hspace{-5mm} Lippincott-Schwartz, J., Donaldson, J. G.,
Schweizer, A., Berger, A., Hauri, E. G., Yuan, L. C.,
and Klausner, R. D. 1990. {\it Cell}  60: 821-836.

\vspace{4mm}
\noindent \hspace{-5mm} Marcus, R. 1955. Calculation of 
Thermodynamic Properties of Polyelectrolytes. 
{\it J. Chem. Phys.} 23: 1057-1068.

\vspace{4mm}
\noindent \hspace{-5mm} Mui, B. L.-S., D\"{o}bereiner, H.-G.,
Madden, T. D., and Cullis, P. R. 1995. 
Influence of Transbilayer Area Asymmetry on the
Morphology of Large Unilamellar Vesicles. 
{\it Biophys. J.} 69: 930-941.

\vspace{4mm}
\noindent \hspace{-5mm} Mutz, M., and Helfrich, W. 1990. Bending
rigidities of some biological model membranes as obtained
from the Fourier analysis of contour sections. 
{\it J. Phys. France.}  51: 991-1002.

\vspace{4mm}   
\noindent \hspace{-5mm} Mitchell, D. J., and Ninham, B. W. 1989.
Curvature Elasticity of Charged Membranes. {\it Lamgmuir}
5: 1121-1123.

\vspace{4mm}   
\noindent \hspace{-5mm} Petrov A. G. and Bivas, I. 1984. {\it
Prog. Surf. Sci.} 16: 389-512.


\vspace{4mm}
\noindent \hspace{-5mm} Rambourg A. and Clermont Y. 1990.
Three-dimensional electron microscopy:
structure of the Golgi apparatus, {\it Eur. 
J. Cell Biology.} 51: 189-200.

\vspace{4mm}
\noindent \hspace{-5mm} Rothman, J. E.
1994. Mechanisms of intracellular protein transport, 
{\it Nature} 372: 55-63.

\vspace{4mm}
\noindent \hspace{-5mm} Sack, F. D., Priestley, D. A., and 
Leopold, A. C. 1983. Surface charge on isolated
maize-coleoptile amyloplasts. {\it Planta}  157: 511-517.

\vspace{4mm}
\noindent \hspace{-5mm} Schekman, R. and Orci, L., 1996.
Coat proteins and vesicle budding,
{\it Science} 271: 1526-1533.

\vspace{4mm}
\noindent \hspace{-5mm} Seifert, U. and Lipowsky, R. 1995. 
Morphology of Vesicles. {\it In} Handbook 
of Biological Physics, v. 1. Structure and
Dynamics of Membranes, Generic and Specific
Interactions. R. Lipowsky and E. Sackmann, editors.
Elsevier Science Publishers B.V.: Amsterdam. 403-464.

\vspace{4mm}
\noindent \hspace{-5mm} Sharp, K. A. and Honig, B. 1990.
Calculating Total Electrostatic Energies with the Nonlinear
Poisson-Boltzmann Equation. {\it J. Phys. Chem.}
94: 7684-7692.

\vspace{4mm}
\noindent \hspace{-5mm} Shraiman, B. I. 1996. 
private communication

\vspace{4mm}
\noindent \hspace{-5mm} Song, J. and Waugh, R. E. 1990. Bending
rigidity of SOPC membranes containing cholesterol. 
{\it J. Biomech. Eng.} 112: 235-240.

\vspace{4mm}
\noindent \hspace{-5mm} Song, J. and Waugh, R. E. 1993. Bending
rigidity of SOPC membranes containing cholesterol. {\it
Biophys. J.}  64: 1967-1970.

\vspace{4mm}
\noindent \hspace{-5mm} Tocanne, J.-F. and Teissi\'{e}, J. 1990.
Ionization of phospholipids and 
phospholipid-supported interfacial
lateral diffusion of protons in membrane model systems.
{\it Biochim. et Biophys. Acta.}  1031: 111-142.

\vspace{4mm}
\noindent \hspace{-5mm} Terasaki, M., Chen, L. B., and
Fujiwara, K. 1986. Microtubules and the 
Endoplasmic Reticulum are Highly
Interdependent Structures. {\it J. Cell Bio.} 
103: 1557-1568.

\vspace{4mm}
\noindent \hspace{-5mm} Trowbridge, I. S., Collwan, J. F.,  
and Hopkins, C. R., 1993. Signal-dependent
membrane protein trafficking in the endocytic pathway,
{\it Ann. Review of Cell Biology}.  9: 129-161.

\vspace{4mm}
\noindent \hspace{-5mm} Winterhalter, M. and Helfrich, W. 1988.
Effect of Surface Charge on the Curvature Elasticity
of Membranes.  {\it J. Phys. Chem.}  92: 6865-6867.

\vspace{4mm}
\noindent \hspace{-5mm} Winterhalter, M. and Helfrich, W. 1992.
Bending Elasticity of Electrically Charged Bilayers:Coupled
Monolayers, Neutral Surfaces, and Balancing Stresses.
{\it J. Phys. Chem.}  96: 327-330.

\end{document}